\documentclass[showpacs, pra, onecolumn, preprintnumbers,   superscriptaddress, aps]{revtex4-2}
\usepackage{eurosym}
\usepackage{color}
\usepackage{amsmath, amssymb}
\usepackage{pifont}
\usepackage{bbold}
\usepackage{float}
\usepackage{tikz}
\usepackage{graphicx}
\usepackage{dcolumn}
\usepackage{bm}
\usepackage{multirow}
\usepackage[caption=false]{subfig}
\usepackage[colorlinks]{hyperref}
\usepackage{placeins}
\usepackage{physics}
\usepackage{mathtools}
\usepackage{amsmath}
\usepackage{amsfonts}
\usepackage{bm}
\makeatletter
\renewcommand\@fnsymbol[1]{*} 
\makeatother
\graphicspath{{Figures/}}

\DeclareMathAlphabet{\bi}{OML}{cmm}{b}{it}
\def \be{\begin{equation}}
\def \ee{\end{equation}}
\def \bearr{\begin{eqnarray}}
\def \eearr{\end{eqnarray}}

\begin{document}

\title{Optically Controlled Topological Phases in the Deformed $\alpha $-$%
T_{3}$ Lattice}
\date{\today}
\author{O.~Benhaida}
\email{othmane.benhaida@gmail.com}
\affiliation{LPHE, Modeling and Simulations, Faculty of Science, Mohammed V University in Rabat, Rabat, Morocco}
\affiliation{CPM, Centre of Physics and Mathematics, Faculty of Science, Mohammed V University in Rabat, Rabat, Morocco}

\author{E.~H.~Saidi}
\affiliation{LPHE, Modeling and Simulations, Faculty of Science, Mohammed V University in Rabat, Rabat, Morocco}
\affiliation{CPM, Centre of Physics and Mathematics, Faculty of Science, Mohammed V University in Rabat, Rabat, Morocco}
\affiliation{College of Physical and Chemical Sciences, Hassan II Academy of Sciences and Technology, Rabat, Morocco}

\author{L.~B.~Drissi}
\email[Corresponding author: ]{ldrissi@fsr.ac.ma} 
\affiliation{LPHE, Modeling and Simulations, Faculty of Science, Mohammed V University in Rabat, Rabat, Morocco}
\affiliation{CPM, Centre of Physics and Mathematics, Faculty of Science, Mohammed V University in Rabat, Rabat, Morocco}
\affiliation{College of Physical and Chemical Sciences, Hassan II Academy of Sciences and Technology, Rabat, Morocco}

\begin{abstract}
Haldane's tight-binding model, which describes a Chern insulator in a
two-dimensional hexagonal lattice, exhibits quantum Hall conductivity
without an external magnetic field. Here, we explore an $\alpha -T_{3}$
lattice subjected to circularly polarized off-resonance light. This lattice,
composed of two sublattices (A and B) and a central site (C) per unit cell,
undergoes deformation by varying the hopping parameter $\gamma _{1}$ while
keeping $\gamma _{2}$= $\gamma _{3}$= $\gamma $. Analytical expressions for
quasi-energies in the first Brillouin zone reveal significant effects of
symmetry breaking. Circularly polarized light lifts the degeneracy of Dirac
points, shifting the cones from M. This deformation evolves with $\gamma _{1}
$, breaking symmetry at $\gamma _{1}=2\gamma $, as observed in Berry
curvature diagrams. In the standard case ($\gamma _{1}=\gamma $),
particle-hole and inversion symmetries are preserved for $\alpha =0$ and $%
\alpha =1$. The system transitions from a semi-metal to a Chern insulator,
with band-specific Chern numbers: $C_{2}=1$, $C_{1}=0$, and $C_{0}=-1$ for $%
\alpha <1/\sqrt{2},$ shifting to $C_{2}=2$, $C_{1}=0$, and $C_{0}=-2$ when $%
\alpha \geqslant 1/\sqrt{2}.$For $\gamma _{1}>2\gamma $, the system enters a
trivial insulating phase. These transitions, confirmed via Wannier charge
centers, are accompanied by a diminishing Hall conductivity. Our findings
highlight tunable topological phases in $\alpha -T_{3}$ lattices, driven by
light and structural deformation, with promising implications for quantum
materials.

\end{abstract}

\maketitle

\section{Introduction}
Higher-order topological phases have revolutionized our understanding of
quantum materials \cite{124,125}, by extending the concept of topology
beyond conventional edge or surface states to boundary modes localized at
corners or hinges \cite{S1 16}-\cite{114}. These phases represent a profound
advancement in the study of topological phenomena, challenging traditional
classifications based on symmetry and topology \cite{121,122}. This
framework builds upon earlier discoveries such as the quantum Hall effect 
\textbf{\cite{In1,In2}}, where quantized Hall conductivity is governed by
topological invariants like the Chern number \textbf{\cite{In5,In4}}.
Haldane's seminal work on breaking time-reversal symmetry in honeycomb
lattices provided a theoretical foundation for understanding these
invariants and their role in phase transitions \textbf{\cite{In9}},
inspiring experimental realizations in diverse geometries and materials
including Lieb and Kagome lattices \cite{In12,In14}, iron-based honeycomb
ferromagnetic insulators \cite{In17}, and electronic and photonic systems 
\cite{In18,In19,Ins18,Ins19}. 

Two-dimensional topological insulators are defined by time-reversal symmetry
and a $\mathbb{Z}_{2}$ topological invariant \cite{In21,In22}. Breaking this
symmetry triggers a topological phase transition, forming a Chern insulator
characterized by chiral edge states and the quantized Hall effect \cite{In25}%
. The Chern number determines the system's topological phase, with zero
indicating a trivial phase. Additionally, circularly polarized off-resonant
light can induce topological transitions \cite{In26}, driving systems from
trivial to non-trivial phases through Floquet theory \cite{In28,In29}.
Second-order photon processes allow for band gap tuning, enabling effective
Hamiltonians dependent on light frequency and intensity, as in the Haldane
model. This framework explains transitions like the polarized light-induced
transition of semi-metallic graphene into a Chern insulator \cite{In30}.
Such prominent example under non-resonant polarized light involves a
phenomenon corroborated by numerous analogous experiments on radiative
systems \cite{In33,In34}.

While most Chern insulators have a Chern number of 1, research is
increasingly focused on higher Chern numbers, which have been observed
experimentally in systems like {thin film magnetic topological insulators
and photonic crystals}\  \cite{In36}. It have been also predicted
theoretically in the Dirac-Weyl semimetals on the $\alpha -T_{3}$ lattice 
\cite{In37}. Numerous studies have revealed remarkable properties of the $%
\alpha -T_{3}$ system, including enhanced Hall conductivity and
unconventional Berry phase effects, which are closely tied to its unique
electronic structure \cite{In38,In40}. For instance, the $\alpha -T_{3}$
lattice, characterized by a tunable parameter controlling the weight of its
flat band, exhibits intriguing phenomena like Klein tunneling and
Fabry-Perot resonances, which have been extensively studied in both
single-layer and bilayer configurations \cite{In38,In40}. Additionally, the
interplay between the flat band and the Berry phase in these systems leads
to novel magneto-optical properties and phase transitions, such as the
quantum spin Hall phase transition \cite{In42}. Further insights have been
gained from detailed analyses of Floquet states in optically driven $\alpha
-T_{3}$ lattices. Under resonant and circularly polarized light irradiation,
these systems exhibit the opening of Berry phase-dependent optical gaps,
revealing unique topological signatures and symmetry-driven phenomena \cite%
{In43}. Moreover, the dice lattice, a special case of the $\alpha -T_{3}$
model with a flat band parameter, has shown to exhibit higher Chern numbers,
accompanied by a significant enhancement in unconventional Hall conductivity 
\cite{In46}.

High Chern numbers have also been reported in decorated lattices,
multi-orbital systems, and lattices with spin-orbit coupling or ultracold
gases, broadening the understanding of topological materials. Introducing
the Haldane model into a bilayer of the $\alpha -T_{3}$ lattice has resulted
in observed Chern numbers of up to 5 \cite{In47}, along with a substantial
enhancement of the $6e^{2}/h$ Hall conductivity \cite{In47,M3}. Other
lattice structures, such as the decorated honeycomb or starlike lattices 
\cite{In48}, as well as a high number of Chern in the Kitaev model in the presence of a high magnetic field applied to the star lattice \cite{IntrR22}, and multi-orbital triangular lattices \cite{In49}, also exhibit
high Chern numbers and larger jumps. High Chern numbers have also been
reported in Dirac \cite{In50} and semi-Dirac \cite{In51} systems.
Additionally, the presence of spin-orbit coupling in honeycomb lattices \cite%
{In52} and ultracold gases in triangular lattices \cite{In54} has been shown
to give rise to high Chern numbers. These studies contribute to a deeper
understanding of the topological properties of these advanced materials.%
\newline

In this work, we explore the topological properties of the $\alpha -T_{3}$
lattice, a fascinating system that generalizes the honeycomb lattice. The $%
\alpha -T_{3}$ lattice consists of two sublattices, A and B, representing
carbon atoms located at the vertices of a hexagonal structure, and a third
site, C, positioned at the center of each hexagon, as illustrated in Fig. %
\ref{Fig0}. Within this structure, quasiparticles can hop from site C to
site A within the same hexagon. The hopping energy between sites A and B is
described as $\gamma \cos \phi $, while that between sites A and C is $%
\gamma \sin \phi $. The parameter\ $\alpha $, defined by $\tan \phi =\alpha $%
, governs the lattice configuration. By varying $\alpha $ between 0 and 1,
different lattice structures can be realized: when $\phi =0$ $(\alpha =0)$,
the lattice corresponds to graphene, and when $\phi =\pi /4$ $(\alpha =1)$,
it becomes the dice lattice. Further details about this lattice structure
are provided in \cite{M1}.

In this study, we investigate the topological properties of the $\alpha
-T_{3}$ lattice under the influence of an external off-resonant electric
field. This field induces a term analogous to the Haldane model, breaking
time-reversal symmetry and modulating the band gap, which is pivotal in
determining the system's topological properties. The simultaneous tunability
of the $\alpha $ parameter and the presence of a flat band create a
distinctive energy dispersion. Additionally, we consider the effect of
anisotropic deformations by altering the hopping energies of quasiparticles
between nearest-neighbor bonds. Specifically, the hopping energy along the
position vector $\delta _{1}$ changes to $\gamma _{1}\sin \phi $ for sites A
and C, and $\gamma _{1}\cos \phi $ for sites A and B, while the hopping
energies along vectors $\delta _{2}$ and $\delta _{3}$ remain unchanged. As
the parameter $\gamma _{1}$ increases, the Dirac points approach each other
and eventually merge at the M point in the Brillouin zone when $\gamma
_{1}=2\gamma $, leading to band closure and the loss of topological
properties. This critical point marks a topological phase transition. Beyond
this point, the energy dispersion acquires a Dirac-like form along the $%
k_{x} $-axis, a phenomenon that has been previously observed in graphene 
\cite{In55}, graphene bilayers \cite{In56}, and the dice lattice \cite{In57}%
. Furthermore, this method has been applied to the single-layer honeycomb
structure $Si_{2}O$, yielding a semi-Dirac dispersion \cite{In58}.

To gain deeper insights, we study the effects of deformation on the
topological properties by tuning the hopping energy $\gamma _{1}$ to
specific values. We calculate the Berry curvature, which encapsulates key
symmetries such as particle-hole symmetry, inversion symmetry, and $C_{3}$
symmetry in the standard case ($\gamma _{1}=\gamma $). As $\gamma _{1}$
deviates from $\gamma $, the $C_{3}$ symmetry is broken, leading to
topological transitions. We also compute the Chern number, which indicates
the system's transition from a non-trivial to a trivial topological phase.
To further confirm this transition, we analyze the evolution of the Wannier
charge center along a closed loop in the Brillouin zone, representing the
average charge position within the unit cell. This analysis is consistent
with the behavior of surface energy bands \cite{M9,M10,M11}. Moreover, we
investigate the quantum Hall conductivity and its response to the
deformation of the energy spectrum as $\gamma _{1}$ varies. Our results
provide valuable insights into the interplay between lattice deformation and
topological phase transitions, shedding light on the tunable nature of
topological properties in $\alpha -T_{3}$ lattices.\newline
The originality of the present work lies in its comprehensive analysis of
the interplay between lattice deformation, external electric fields, and
topological properties in the $\alpha -T_{3}$ lattice. By exploring the
combined effects of anisotropic hopping energies, Berry curvature, and
quantum Hall conductivity, this study not only advances the understanding of
topological materials but also paves the way for potential applications in
quantum technologies.

This article is organized as follows: Section \ref{secII} presents the
Hamiltonian describing quasiparticle dynamics in irradiated and deformed $%
\alpha -T_{3}$ lattices. Section \ref{secIII} analyzes and discusses the
quasi-energy spectrum. Section \ref{secIV} investigates the topological
properties of this lattice, including the Berry curvature (Subsection \ref%
{SubsecA}), the Chern phase diagram (Subsection \ref{SubsecB}), and the
evolution of the Wannier charge center (Subsection \ref{SubsecC}). Section %
\ref{secV} examines the anomalous Hall conductivity. Finally, Section \ref%
{secVI} provides a conclusion and summary of the work.

\section{Model and Hamiltonian}\label{secII}

\begin{figure}[H]
\centering
\includegraphics[width=0.6\linewidth]{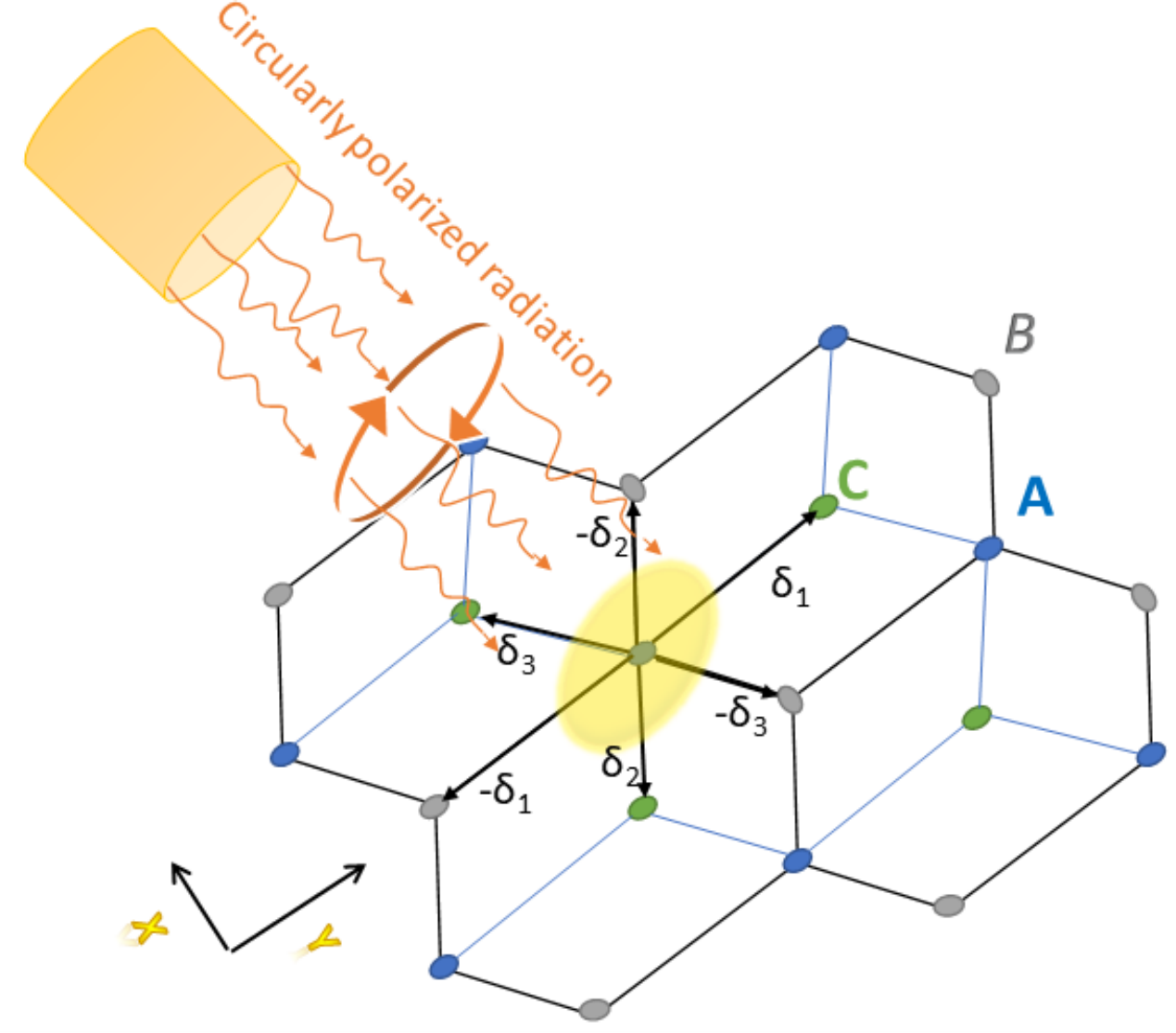}
\caption{Schematic of a deformed $\protect \alpha -T_{3}$ lattice exposed to
circularly polarized off-resonance light, where the deformation affects only
the position vector $\protect \delta _{1}$, changing the hopping energy from $%
\protect \gamma _{1}=\protect \gamma $ to 3$\protect \gamma $, while the
hopping energies $\protect \gamma _{1}=\protect \gamma $ and $\protect \gamma %
_{2}=\protect \gamma $ associated with the position vectors $\protect \delta %
_{2}$ and $\protect \delta _{3}$ remain unchanged.}
\label{Fig0}
\end{figure}

The rescaled tight-binding Hamiltonian \cite{M1} that describes the motion of a quasiparticle along a $p_{z}$ orbital within an $\alpha -T_{3}$ lattice, between its nearest neighbours within the basis sublattices B, A, and C, is given by:. 
\begin{equation}
H(\bm k)=%
\begin{pmatrix}
0 & \cos (\varphi )\rho (\bm k) & 0 \\ 
\cos (\varphi )\rho ^{\ast }(\bm k) & 0 & \sin (\varphi )\rho (\bm k) \\ 
0 & \sin (\varphi )\rho ^{\ast }(\bm k) & 0%
\end{pmatrix}%
\end{equation}%
where $\rho (\bm k)=\sum_{j=1}^{3}\gamma _{j}e^{i\bm k\bm{\delta_{i}}}$, and $\bm{\delta_{j}}$ denotes the vectors connecting the nearest neighbour sites. These vectors are given by: $\bm \delta _{1}=a(0,1)$,  $\bm \delta _{2}=a(-\sqrt{3}/2,-1/2)$ and $\bm\delta _{3}=a(\sqrt{3}/2,-1/2)$, where $a$ represents the distance between two nearest neighbours, as illustrated in Fig.\ref{Fig0}. The hopping energy between sites A and B is given by $\gamma_{j} \cos(\varphi)$, and the hopping energy between sites A and C is $\gamma_{j}\sin(\varphi)$. The parameter $\alpha$, defined as $\tan(\varphi)=\alpha$, governs the structure of the $\alpha-T_{3}$ lattice. This parameter is bounded within the range $0\leqslant\alpha \leqslant 1$, with $\alpha=0$ corresponding to the graphene lattice and $\alpha=1$ to the dice lattice. In our study, we introduce a uniaxial deformation of the $\alpha-T_{3}$ lattice as a theoretical tool to explore topological transitions. A change models this deformation in the hopping amplitudes between neighbouring sites in the $\bm \delta _{1}$ and -$\bm \delta _{1}$ directions, which are associated with a hopping amplitude $\gamma_{1}$ representing the effect of the deformation. In contrast, the amplitudes in the other directions ($\bm \delta _{2}$, $\bm \delta _{3}$, -$\bm \delta _{2}$, -$\bm \delta _{3}$) are kept constant and equal to $\gamma_{2}=\gamma_{3}=\gamma$. This approach simulates externally imposed anisotropy, such as uniaxial pressure, rather than deformation induced dynamically by an optical field. The relevance of this model is based on previous work that introduced uniaxial modifications of hopping terms to study topological phases\cite{In55,In56,In57}. These studies confirm that the increase in overlap between atomic orbitals, induced by uniaxial deformation, leads to a growth in hopping amplitudes. Thus, the strain introduced in our model allows fine modulation of topological phase transitions, particularly when combined with an off-resonance light field. In this section, we'll study photon-electron interactions by applying a polarised electric field perpendicular to the plane of the deformed $\alpha-T_{3}$ lattice, as illustrated in Fig.\ref{Fig0}. The electric field is derived from the vector potential with respect to time $\partial _{t}A(t)$, and this vector potential depends on time, $\bm A(t)=A_{0}\left( \cos (\omega t),\sin (\omega t)\right) $, where $\omega $ and $A_{0}$ are the radiation frequency and the vector potential amplitude, respectively.\newline
when an electric field is applied to electrons moving from site $\bm m$ to
the nearest site $\bm m+\mathbf{\delta }$, they gain energy, represented by
the hopping energy $\gamma _{j}$, which is transformed into $\gamma
_{j}e^{i\varpi }$, where $\varpi =e/\hbar \int_{\bm m}^{\mathbf{m}+\bm \delta
}\bm A(t)d\mathbf{x}$ is the phase factor gained by the electron. The
Hamiltonian then becomes 
\begin{equation}
H(\bm k,t)=%
\begin{pmatrix}
0 & \cos (\varphi )\rho (\bm k,t) & 0 \\ 
\cos (\varphi )\rho ^{\ast }(\bm k,t) & 0 & \sin (\varphi )\rho (\bm k,t) \\ 
0 & \sin (\varphi )\rho ^{\ast }(\bm k,t) & 0%
\end{pmatrix}%
,
\end{equation}%
Where $\rho (\bm k,t)=\sum_{j=1}^{3}\gamma _{j}e^{i(\mathbf{\bm k}+e\mathbf{A}(t))\mathbf{\delta _{j}}}=\left[ \gamma_{1}e^{i\varsigma \sin(wt)}e^{i\bm k\mathbf{\delta }_{1}}+\gamma e^{i\bm k\bm \delta_{2}}e^{-i\varsigma \sin(\pi/3+\omega t)}+\gamma e^{i\bm k\bm \delta _{3}}e^{i\varsigma \sin(\pi/3-\omega t)}\right] $ and we have a periodic 
Hamiltonian in time, such that $H(\bm k, t +T) = H(\bm k,t)$, 
with $T=2\pi/\omega$. This periodicity enables the Hamiltonian 
to be expressed as a Fourier series: $H(\bm k, t ) =\sum_{n=-\infty}^{+\infty} H_n e^{in\omega t}$. In this framework, a factor in  $\rho (\bm k,t)$ arising from light-matter coupling contains an exponential of a trigonometric function oscillating with respect to t, containing $e^{i\varsigma \sin(\omega t)}$, $e^{-i\varsigma \sin(\pi/3+\omega t)}$ and $e^{i\varsigma \sin(\pi/3-\omega t)}$. We can therefore apply the Jacobi-Anger 
expansion, $e^{i\chi\sin\beta} =\sum_{n=-\infty}^{+\infty} J_n(\chi) 
e^{in\beta}$, where $J_n(\chi)$ is the nth-order Bessel function 
of the first kind. This expansion explicitly identifies the 
Fourier components of $H(\bm k, t)$, denoted $H_n$. We can therefore rewrite $\rho (\bm k,t)$ as $\rho (\bm k,t)=\sum_{n=-\infty}^{\infty }J_{n}(\varsigma )\left[ \gamma_{1}e^{inwt}e^{i\bm k\mathbf{\delta }_{1}}+\gamma e^{i\bm k\bm \delta_{2}}e^{-in(wt+\pi /3)}+\gamma e^{i\bm k\bm \delta _{3}}e^{in(\pi /3-wt)}\right] $. The intensity of
light is characterized by a dimensionless parameter $\varsigma
=eA_{0}a/\hbar $ and is much smaller than 1 when the laser frequency is sufficiently high, where
$e$ is the electric charge. We consider that the light incident on the $\alpha -T_{3}$ lattice has an
off-resonance frequency, as explained in \cite{In26}. When an electron is
subjected to an off-resonance frequency, it does not excite it directly but
simply modifies the band structure in an effective way by absorbing and
emitting virtual photons. An off-resonance state is reached when the
frequency of the photons is well above the bandwidth, i.e. $\omega >>3\gamma
_{1}$. We have a time-dependent and periodic Hamiltonian $H(T+t,k)=H(t,k)$.
Floquet's theorem is perhaps the most convenient solution to this problem,
with $T=2\pi /\omega $. In this context, the properties of the interaction between light and matter in an off-resonant system can be described by an effective Hamiltonian derived from Floquet theory in the high-frequency limit \cite{effctiv1,effctiv2,effctiv3,M3}, which is given by the following formula: 
\begin{equation}
H_{eff}(\bm k)=H_{0}(\bm k)+\sum_{s=1}^{+\infty}\dfrac{1}{s\hbar \omega }\left[ H_{-s}(\bm %
k),H_{+s}(\bm k)\right] +\epsilon (1/\omega ^{2}).
\end{equation}%
In this framework, couplings between replicas that differ by more than one photon quantum are considered negligible. Indeed, at low energies, the Fourier components $H_s$ annul for any $|s| >1$. Consequently, the analysis only considers the components corresponding to $|s| \leqslant1$. In this case, we can find an effective Hamiltonian that is independent of time.
\begin{equation}
H_{eff}(\bm k)=H_{0}(\bm k)+\dfrac{1}{\hbar \omega }\left[ H_{-1}(\bm %
k),H_{+1}(\bm k)\right] +\epsilon (1/\omega ^{2}).
\end{equation}%

 The effective Hamiltonian was also applied to the undeformed $\alpha-T_{3}$ lattice\cite{In37,In43}. We express the time-dependent Fourier components of the Hamiltonian as
follows: 
\begin{equation}
H_{s}(\bm k)=\dfrac{1}{T}\int_{0}^{T}H(\bm k,t)e^{-iswt}dt.
\end{equation}%
The second term is responsible for the absorption of a virtual photon by an
electron through $H_{1}H_{-1}$ and its emission through $H_{-1}H_{1}$. We
give the explicit expression of the effective Hamiltonian considering terms
up to $\epsilon (1/\omega )$. 
\begin{equation}
H_{eff}(\bm k)=%
\begin{pmatrix}
\eta (\bm k)\cos (\varphi )^{2} & J_{0}(\varsigma )\cos (\varphi )\rho (\bm %
k) & 0 \\ 
J_{0}(\varsigma )\cos (\varphi )\rho ^{\ast }(k) & -\eta (\bm k)\cos
(2\varphi ) & J_{0}(\varsigma )\sin (\varphi )\rho (\bm k) \\ 
0 & J_{0}(\varsigma )\sin (\varphi )\rho ^{\ast }(\bm k) & -\eta (\bm k)\sin
(\varphi )^{2}%
\end{pmatrix}%
,
\end{equation}%
and $\eta (\bm k)$ are defined as follows 
\begin{equation}
\eta (\bm k)=2\Delta (-\cos [\dfrac{\sqrt{3}k_{x}a}{2}]+\dfrac{\gamma _{1}}{%
\gamma }\cos (\dfrac{3k_{y}a}{2}))\sin (\dfrac{\sqrt{3}k_{x}a}{2}).
\end{equation}%
where $\Delta =\dfrac{\sqrt{3}\gamma ^{2}\varsigma ^{2}}{2\hbar \omega }$.
We consider a weak conduit $\varsigma <<1$, which implies taking the
approximation $J_{0}(\varsigma )\approx 1$ and $J_{1}(\varsigma )\approx 
\dfrac{\varsigma }{2}$.
We can obtain an effective Hamiltonian describing the undeformed $\alpha-T_{3}$ lattice ($\gamma_{1}=\gamma$) at low energies, in the vicinity of Dirac points $\bm K'(+)$ and $\bm K(-)$, where the function $\rho(\bm q)$ becomes linear, i.e. $\rho(\bm q) =\pm q_x - iq_y$. The term  
$\eta^{\bm K'/\bm K}=\pm\dfrac{e^{2}A^{2}_{0}v^{2}_{f}}{\hbar \omega}$ then becomes equivalent to $\Delta^{\pm}=\pm\dfrac{e^{2}A^{2}_{0}v^{2}_{f}}{\hbar \omega}$. We thus find the same effective Hamiltonian as demonstrated in reference \cite{In43}, where $v_F=3\gamma a/2\hbar$ represents the Fermi velocity.\newline
$\eta (\bm k)$ is the light-induced term, identical
to the near-second Haldane term with $\phi =\pi /2$ and $t_{2}=\Delta $,
which satisfies the Haldane model for graphene and the dice lattice \cite%
{In46,In55}. This term is responsible for breaking the time reversal
symmetry. This results in a gap in the Dirac points. Later we'll discuss the 
$\gamma _{1}$-effect, which is a mesh distortion that changes the location
of this gap.
As explained previously \cite{In37}, if we take $\gamma_{1}=\gamma$, $%
H_{eff}(\bm k)$ satisfies the anticommutation relations when $\alpha=0$ and $%
\alpha=1$.\newline
\begin{equation}  \label{eq7}
\left \lbrace H^{\alpha=0}_{ eff}(\bm k),C^{\alpha=0}\right \rbrace =0,\quad
\left \lbrace H^{\alpha=1}_{ eff}(\bm k),C^{\alpha=1}\right \rbrace =0.
\end{equation}
where $C^{\alpha=0}$ is defined as the graphene operator and $C^{\alpha=1}$
as the dice operator, given as: 
\begin{equation}
C^{\alpha=0}%
\begin{pmatrix}
0 & -1 & 0 \\ 
1 & 0 & 0 \\ 
0 & 0 & 1%
\end{pmatrix}%
\mathcal{K},\quad C^{\alpha=1}%
\begin{pmatrix}
0 & 0 & -1 \\ 
0 & 1 & 0 \\ 
-1 & 0 & 0%
\end{pmatrix}%
\mathcal{K}.
\end{equation}
with $\mathcal{K}$ as the complex conjugate, equation (\ref{eq7}) implies
the existence of a particle and a hole such that $\varepsilon(\bm %
k)=-\varepsilon(-\bm k)$, as well as a flat band at zero. However, when we
take $\gamma_{1}=2\gamma$ and whatever $\alpha$ is, at point $\bm M$ (see
Fig.\ref{fig1}), $H_{eff}(\bm k)$ satisfies the anticommutation relations. 
\begin{equation}
\left \lbrace H_{ eff}(\bm k),C^{\alpha=0}\right \rbrace =0,\quad \left
\lbrace H_{ eff}(\bm k),C^{\alpha=1}\right \rbrace =0.
\end{equation}
In this case, we can conclude that the radiation does not affect the system
in the same way as it would normally, because the system no longer depends
on $\alpha$. We will later demonstrate the importance of modifying $%
\gamma_{1}$, as this results in the deformation of a lattice, a change $%
\gamma_{1}$, and an observation of its effect on topological properties.

\section{Quasienergy and band structure}\label{secIII}
In this section, we study the energy bands as $\gamma _{1}$ varies. The
Hamiltonian will be diagonalized to obtain the eigenvalues. This
diagonalization leads to the characteristic equation, known as the \textbf{%
\textit{depressed cubic equation}}, $\lambda _{3}\varepsilon ^{3}+\lambda
_{1}\varepsilon +\lambda _{0}=0$, whose solutions can be expressed as
follows: 
\begin{equation}
\varepsilon _{\nu }(\bm k)=2\sqrt{\dfrac{-\lambda _{1}}{3}}\cos \left[ 
\dfrac{1}{3}\arccos \left[ \dfrac{3\lambda _{0}}{2\lambda _{1}}\sqrt{\dfrac{%
-3}{\lambda _{1}}}\right] -\dfrac{2\nu \pi }{3}\right] ,
\end{equation}%
and 
\begin{align}
\lambda _{0}& =-\dfrac{1}{8}\eta ^{3}(\bm k)\sin (2\varphi )\sin (4\varphi
),\quad \\
\lambda _{1}& =-\dfrac{1}{8}\left( 8|\rho (\bm k)|^{2}+\eta ^{2}(\bm %
k)(5+3\cos (4\phi ))\right) , \\
\lambda _{3}& =1.
\end{align}%
The quasi-energy associated with the conduction band is represented by the
value $\nu =0$, while the flat band and the valence band are represented by
the values $\nu =1$ and $\nu =2$, respectively. These quasienergies
correspond to normalised pseudo-vectors. 
\begin{equation}
\ket{\Upsilon_{\nu}(\bm k)}=\mathcal{N}_{\nu }(\bm k)%
\begin{pmatrix}
-\frac{\cos (\phi )\rho (\bm k)}{-\varepsilon _{\nu }(\bm k)+\eta (\bm k)} & 
1 & \frac{\sin (\phi )\rho ^{\ast }(\bm k)}{\varepsilon _{\nu }(\bm k)+\eta (%
\bm k)}%
\end{pmatrix}%
^{T},
\end{equation}%
and 
\begin{equation}
\mathcal{N}_{\nu }(\bm k)=\left[ 1+|\rho (\bm k)|^{2}\left( \frac{\cos
^{2}(\phi )}{(-\varepsilon _{\nu }(\bm k)+\eta (\bm k))^{2}}+\frac{\sin
^{2}(\phi )}{(\varepsilon _{\nu }(\bm k)+\eta (\bm k))^{2}}\right) \right]
^{-1/2}.
\end{equation}%
\begin{figure}[H]
	\centering
	\includegraphics[width=0.75\linewidth]{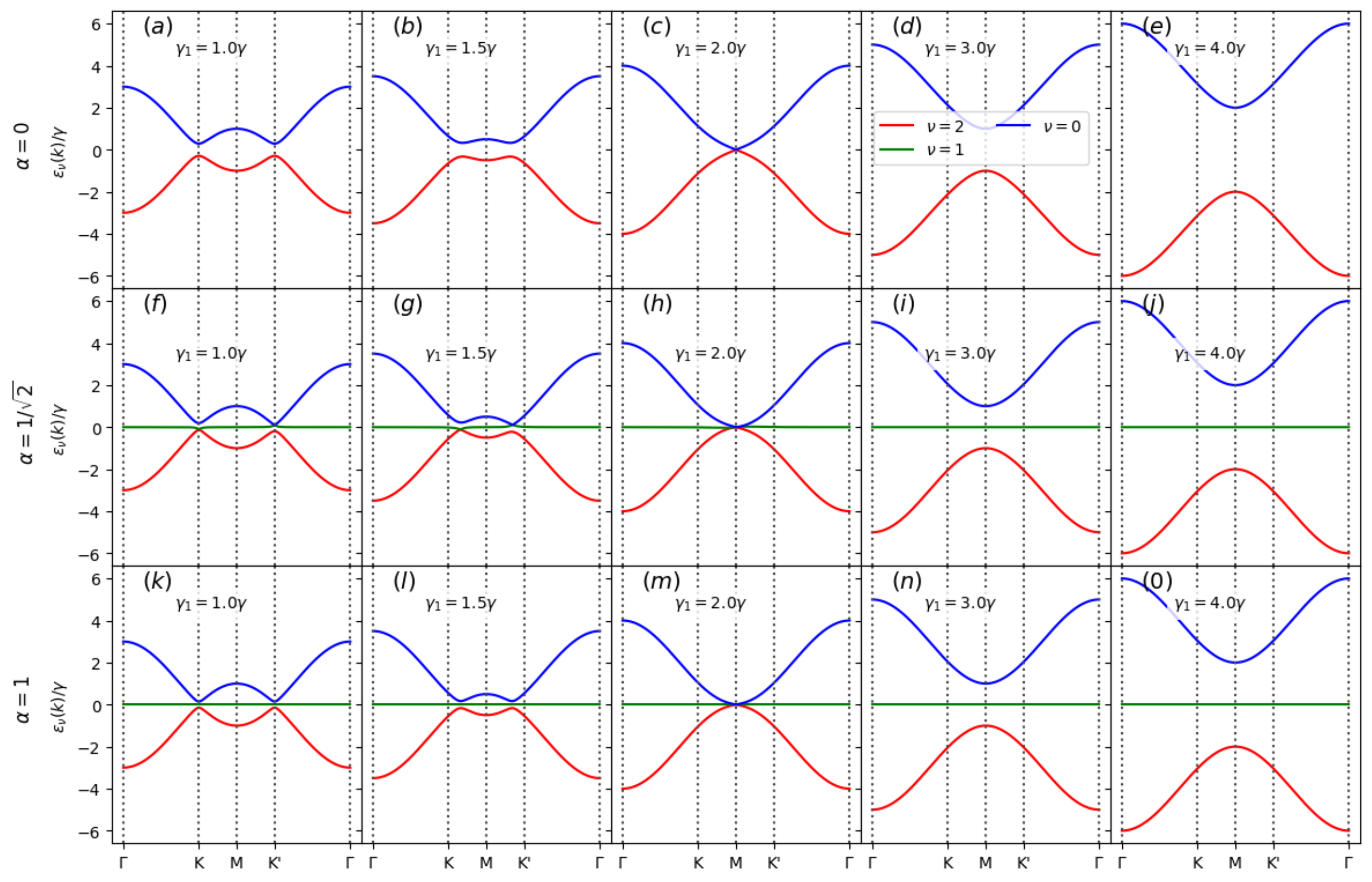}
	\includegraphics[width=0.23\linewidth]{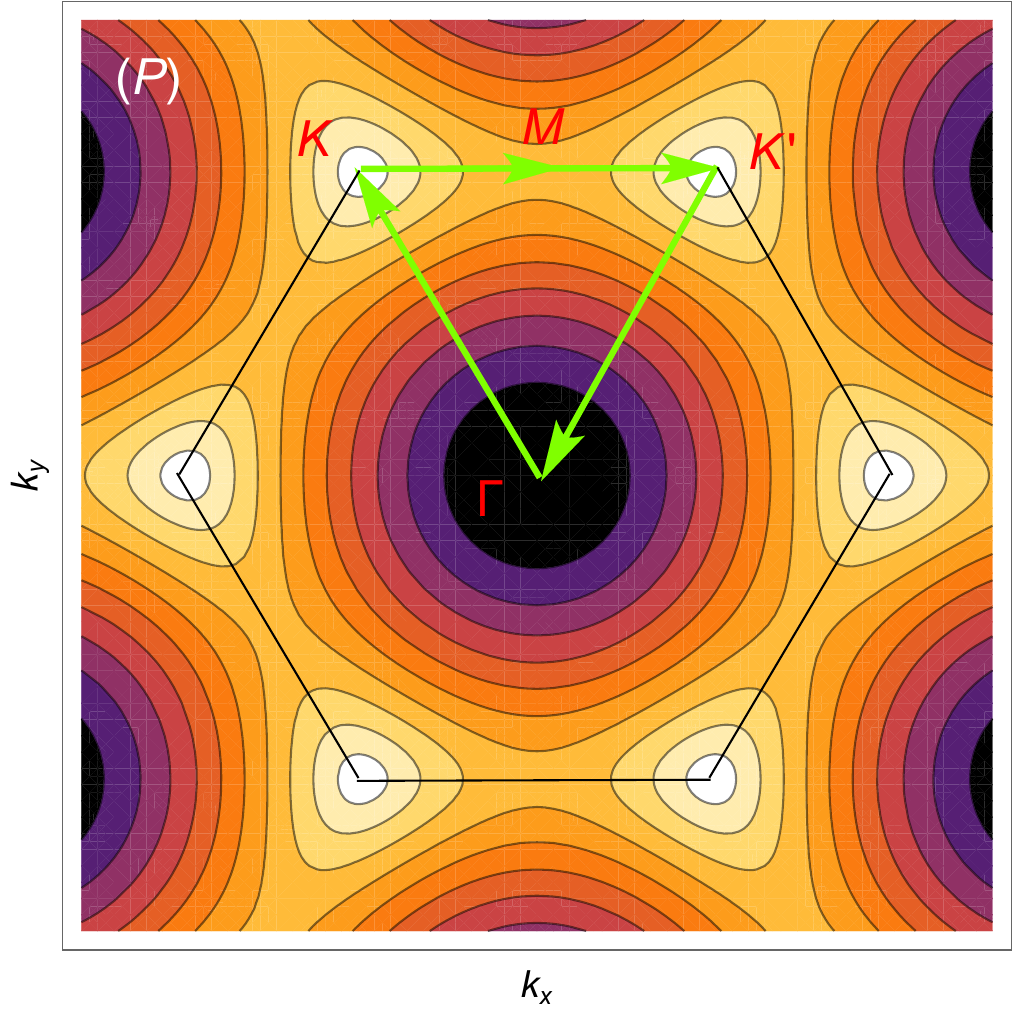}
	\caption{The band structure of the irradiated and deformed $\protect \alpha %
		-T_{3}$ lattice is illustrated as a function of $\protect \gamma _{1}=\protect%
			\beta \protect \gamma $ along the $k_{x}$ axis in the following cases: $%
			\bullet $ (a) $\protect \gamma _{1}=1\protect \gamma $, (b)$\protect \gamma %
			_{1}=1.5\protect \gamma $, (c) $\protect \gamma _{1}=2\protect \gamma $, (d) $%
			\protect \gamma _{1}=3\protect \gamma $,  (e) $%
			\protect \gamma _{1}=4\protect \gamma $  with $\protect \alpha =0$, $\bullet $
			(f) $\protect \gamma _{1}=1\protect \gamma $ , (g) $\protect \gamma _{1}=1.5%
			\protect \gamma $, (h)$\protect \gamma _{1}=2\protect \gamma $, (i) $\protect%
			\gamma _{1}=3 \gamma $, (j) $\protect%
			\gamma _{1}=4 \gamma $ with $\protect \alpha =\dfrac{1}{\protect%
				\sqrt{2}}$, $\bullet $ (k) $\protect \gamma _{1}=1\protect \gamma $, (l) $%
			\protect \gamma _{1}=1.5\protect \gamma $, (m)$\protect \gamma _{1}=2\protect%
			\gamma $, (n) $\protect \gamma _{1}=3\protect \gamma $ ,(o) $\protect \gamma _{1}=4\protect \gamma $ with $\protect \alpha %
			=1$. Sub-figure (P) represents the first Brillouin zone of the hexagonal
			lattice used to calculate and structure the band structure along the path ($%
			\Gamma \rightarrow K\rightarrow M\rightarrow K^{\prime }\rightarrow \Gamma $%
			). We have normalized the Hamiltonian by $\gamma$ to account for this constant in the figures. We used $\Delta/\gamma = 0.27$, with $\hbar \omega = 3\gamma$, and a field intensity amplitude $\varsigma = 0.7$. Taking $\gamma = 2.7$ eV, the frequency range considered is $\omega\in[2.7 PHz, 8.1 PHz]$, which we are considering here within a theoretical framework. This frequency range is physically accessible. Indeed, recent studies have explored controlling electric current by irradiating solid materials at petahertz frequencies \cite{REaddsp1,REaddsp3,REaddsp4}. Furthermore, applications in the fields of ultrafast optical waveforms, digital logic, communications, and quantum computing \cite{REaddsp2,REaddsp3,REaddsp4} make this regime relevant to modern physics.}
	\label{fig1}
\end{figure}
\begin{figure}[H]
	\centering
	\includegraphics[width=0.34\linewidth]{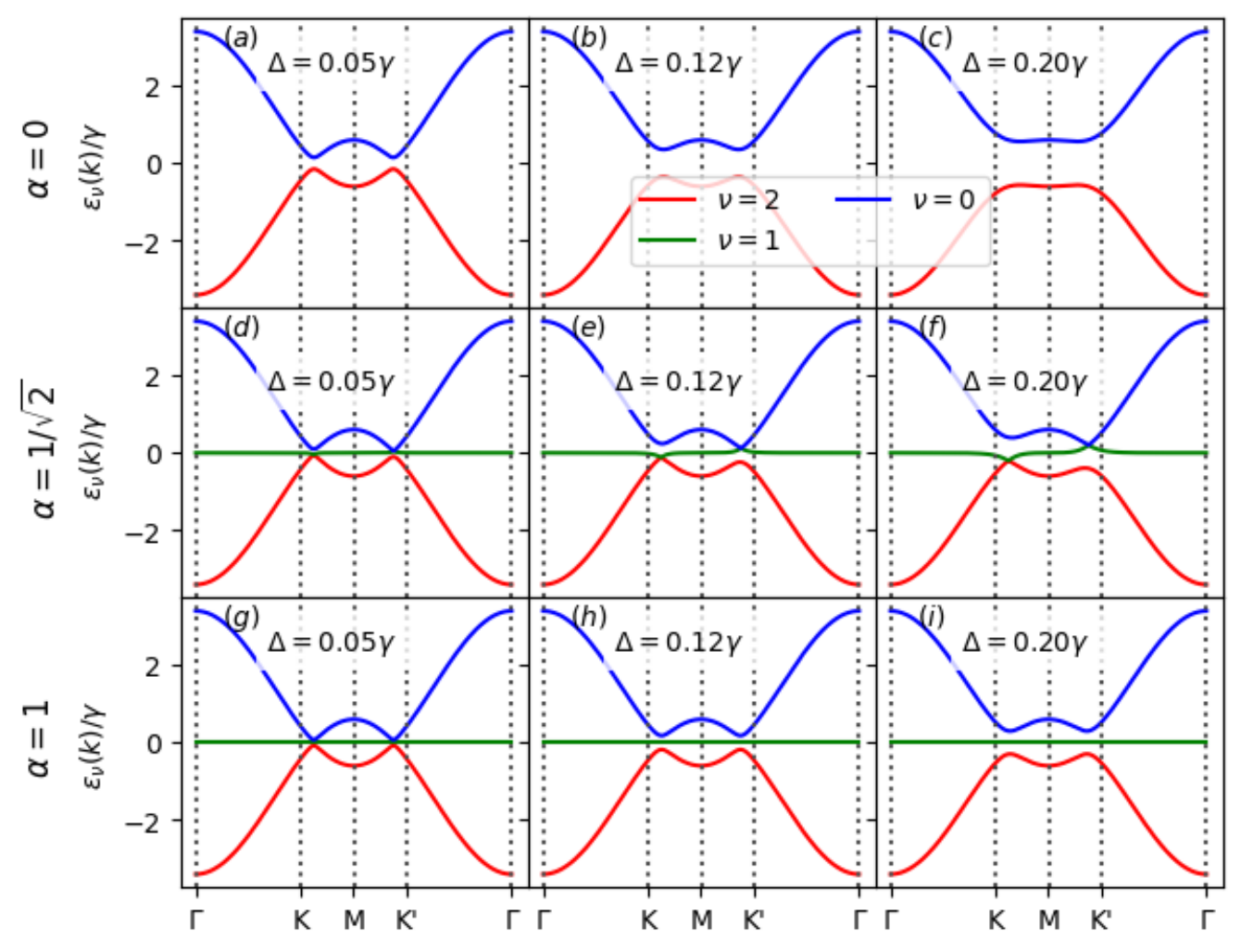}
	\includegraphics[width=0.31\linewidth]{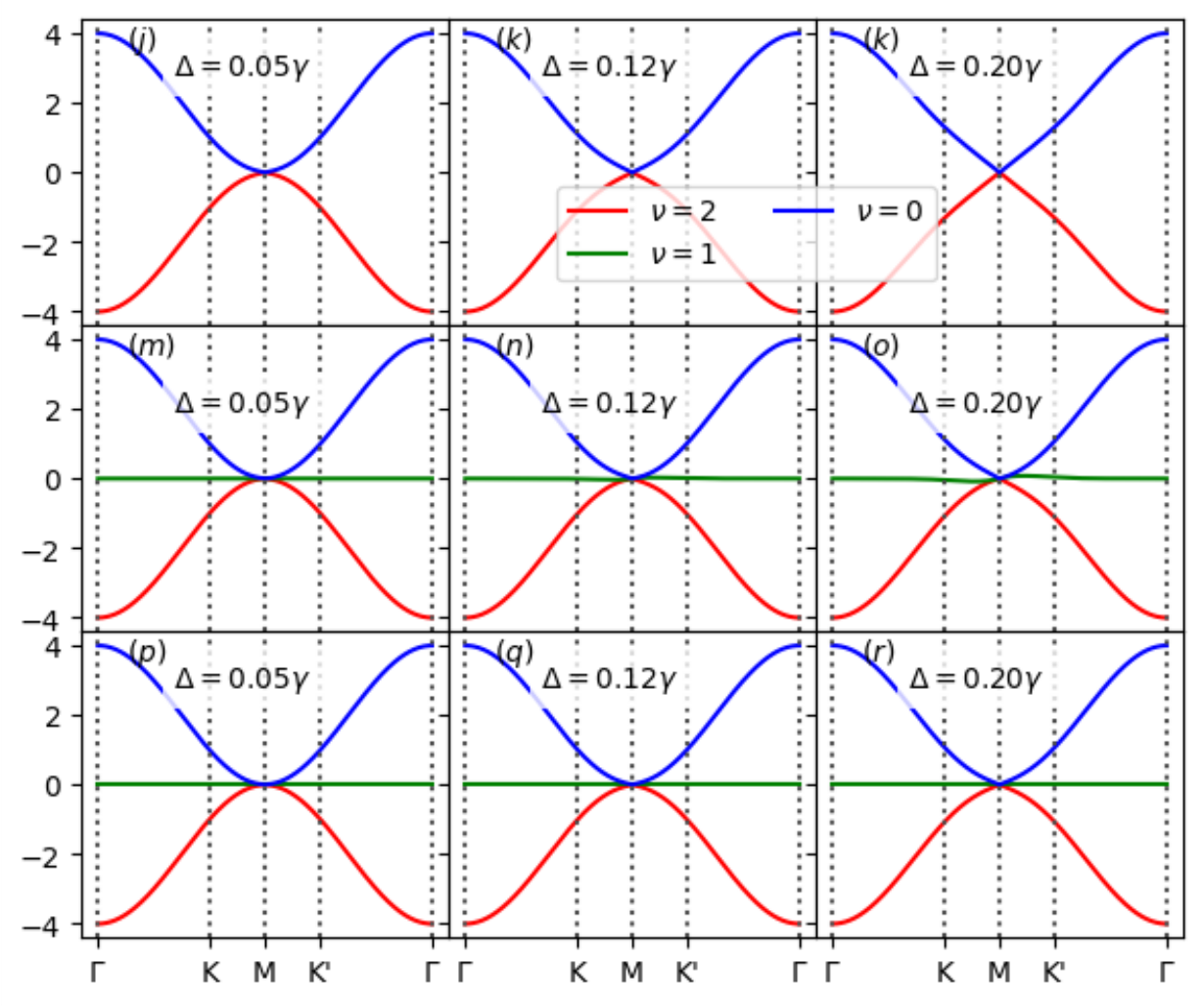}
	\includegraphics[width=0.32\linewidth]{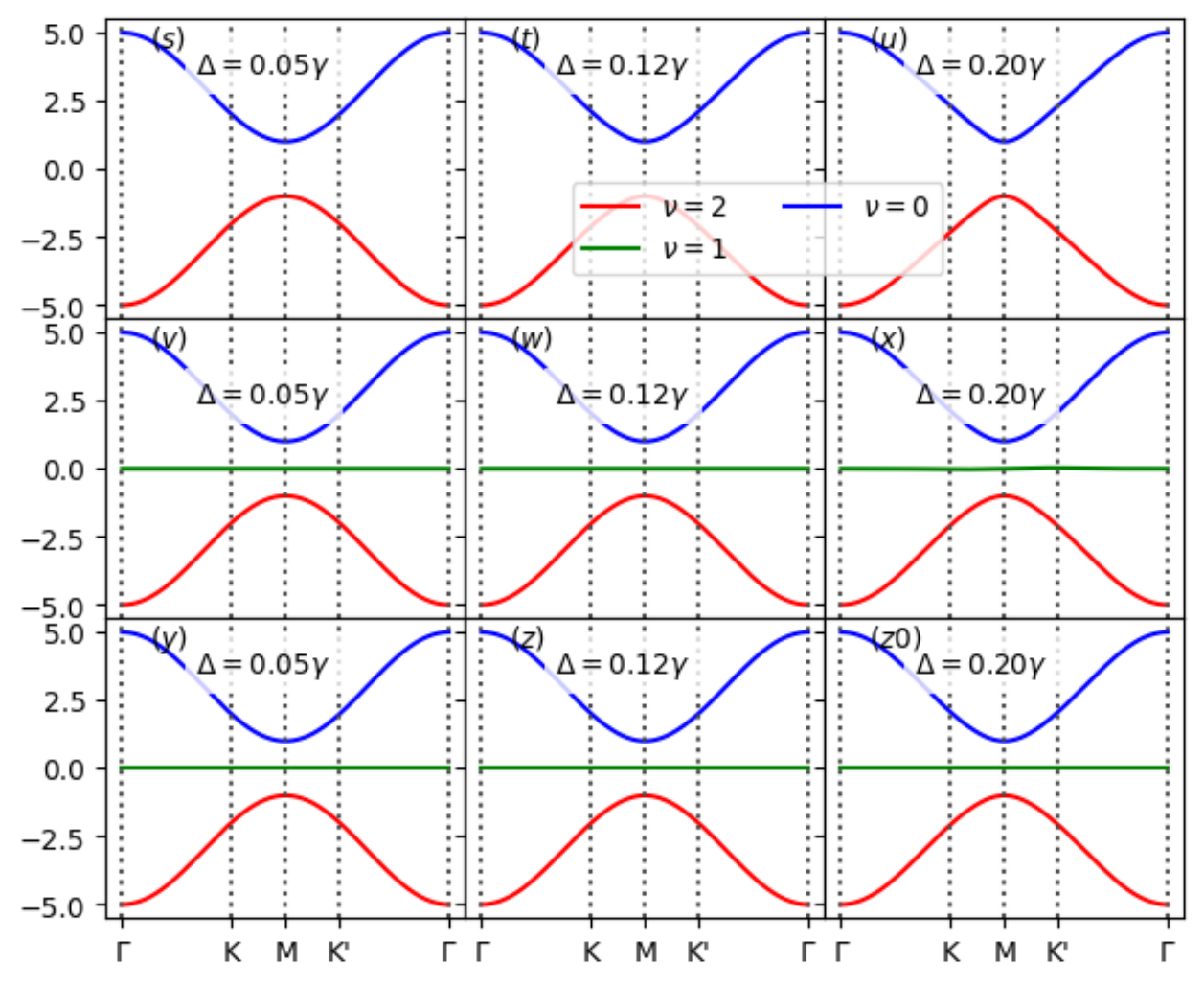}
	\caption{The figure illustrates the band structure in relation to variations in the amplitude $\Delta$. The deformation parameter is controlled by the hopping energy $\gamma_{1}$. For sub-figs (a) to (i), $\gamma_{1}$ is set to 1.4$\gamma$; for sub-figs (j) to (r), it is set to 2$\gamma$; and for subfigs (s) to (z0), it is set to 3$\gamma$. These correspond to the cases $\alpha=0$ for the top plateau, $\alpha=1/\sqrt{2}$ for the middle plateau, and $\alpha=1$ for the bottom plateau. The value of the amplitude $\Delta$ is specified in each sub-fig.}
	\label{fig111}
\end{figure}
Before discussing the band structure, we determine the path of the first
Brillouin zone in the reciprocal lattice, which is in the form of a
hexagonal lattice, to construct the band structure path, as shown in Fig.\ref%
{fig1}-(P). Now, we analyse and discuss the band structure by varying $%
\alpha $ for three values: 0, $1/\sqrt{2}$ and 1. In addition, for each
fixed value of $\alpha $, we vary $\gamma _{1}$ while keeping $\Delta $
constant. The $\alpha -T_{3}$ lattice comprises three bands: the valence band, the flat band and the conduction band. In the absence of irradiation, these bands converge at the Fermi level at the Dirac points $\bm K\left( -\dfrac{2\pi }{3\sqrt{3}a},\dfrac{2\pi }{3a}\right) $ and $\bm  K^{\prime }\left( \dfrac{2\pi }{3\sqrt{3}a},\dfrac{2\pi }{3a}\right) $, resulting in semi-metallic properties for the system. However, when the lattice is deformed and exposed to off-resonance radiation, a Haldane-type term is generated, breaking the time-reversal symmetry and opening a gap. The size of this gap depends on the parameters $\Delta$, $\alpha$ and $\gamma_{1}$. Controlling these parameters and varying $\gamma_{1}$ reveals that the gap progressively decreases in the graphene case ($\alpha=0$), as the system moves away from the Dirac points, as illustrated in subfigures \ref{fig1}(a) to \ref{fig1}(c). This behaviour is similar to that observed in the Haldane model, where the parameters $t_{2} =\Delta$ and $\phi = \pi/2$ are used \cite{In55}. When $\gamma_{1}$ reaches a critical value of $\gamma_{1}=2\gamma_{1}$, the gap closes at point $\bm M$, as shown in Fig. \ref{fig1}(c). For $\gamma_{1}>2\gamma$, the gap reopens, as shown in sub-Figs \ref{fig1}(d) and \ref{fig1}(e).
For $\alpha = 1$, behaviour similar to that observed for $\alpha =0$ is seen, but with an additional flat band present. The gap decreases with increasing $\gamma_{1}$, as illustrated in sub-Figs. \ref{fig1}(k)-\ref{fig1}(m), and closes at point $\bm M$, where the bands become degenerate (see sub-fig. \ref{fig1}(m)). This occurs despite the irradiation-induced breaking of time-reversal symmetry, after which the system exhibits a semi-Dirac-like spectrum. This spectrum is characterised by quadratic dispersion in $k_x$ and linear dispersion in $k_y$ (here, dispersion along $k_y$ is not shown), implying electron propagation anisotropy at $\gamma_{1} = 2\gamma$. Gap opening and closing lead to a phase transition at $\gamma_{1} = 2\gamma$. At $\gamma_{1} > 2\gamma$, the gap opens again, as illustrated in sub-figs \ref{fig1}(n) to \ref{fig1}(o). This behaviour is analogous to that observed in a Dice lattice subjected to a Haldane-type field \cite{In57}.
For $\alpha =1/\sqrt{2}$, the gap partially closes at the Dirac points $\bm K$ and $\bm K'$, as illustrated in Fig. \ref{fig1}(f). This signals a phase transition at $\gamma_{1} = 2\gamma$, as confirmed by references \cite{In37,In43,M3}. Here, the flat band becomes slightly dispersive. As $\gamma_{1}$ increases, another gap partially opens at the $\bm K$ and $\bm K'$ point and narrows until it disappears completely at $\gamma_{1} = 2\gamma$, as shown in sub-figs \ref{fig1}(f)-\ref{fig1}(h), where the bands become fully degenerate. When $\gamma_{1}>2\gamma$, the gap opens again, but the effect of radiation is no longer sufficient to control the adjustment of the gap.\newline
 To better understand the effect of light, we examined the dispersion relationship by fixing $\gamma_{1}$ and varying the amplitude $\Delta$ to observe any effects when $\gamma_{1}>2\gamma$, as illustrated in Fig\ref{fig111}.  We observed that irradiation influences the $\alpha-T_{3}$ lattice by increasing the energy gap for $\alpha=0$ (graphene) and $\alpha=1$ (Dice lattice) as the amplitude  $\Delta$ increases. It also partially widens the gap at the Dirac points $\bm K$ and $\bm K'$ for $\alpha=1/\sqrt{2}$ in the interval $\gamma \leqslant \gamma
_{1}<2\gamma $.  This behaviour can be seen in Figures \ref{fig111}(a) to \ref{fig111}(c) for $\alpha=0$, \ref{fig111}(g) to \ref{fig111}(i) for $\alpha=1$ and \ref{fig111}(d) to \ref{fig111}(f) for$\alpha=1/\sqrt{2}$ with $\gamma_{1}= 1.4\gamma$ fixed.  Conversely, when $\gamma_{1}= 2\gamma$, no noticeable effect of light is observed: the gap closes completely at point $\bm M$, as illustrated in sub-Figs \ref{fig111}(j) to \ref{fig111}(r), regardless of the value of $\alpha$.  For $\gamma_{1} > 2\gamma$, the gap reopens and remains unchanged for all values of $\alpha$, as shown in sub-Figs \ref{fig111}(s) and \ref{fig111}(z0) for $\gamma_{1} =3\gamma$.  In this case, increasing the amplitude $\Delta$ does not affect the opening of the gap. We conclude that the opening and closing of the gap around $\gamma_{1} = 2\gamma$ induces a topological phase transition. The effect of radiation is significant only within the interval $\gamma \leqslant \gamma
_{1}<2\gamma $.\newline
One aspect that merits attention is that, despite the system being subjected
to radiation and $\gamma_{1}$ being fixed at 2$\gamma$, the gap closes and
the time-reversal symmetry remains broken. This could indicate a change in
the topological properties of the $\alpha-T_{3}$ lattice. This is precisely
what we are going to examine in the study of topological properties as $%
\gamma_{1}$ varies.
\section{Topological properties}\label{secIV}
 In this section, we study the effect of deformation from $%
\gamma_{1}$ change on Berry curvature, Chern number, and Wannier charge
center.
\subsection{Berry curvature}\label{SubsecA} 
\begin{figure}[H]
	\centering
	\includegraphics[width=0.8\linewidth]{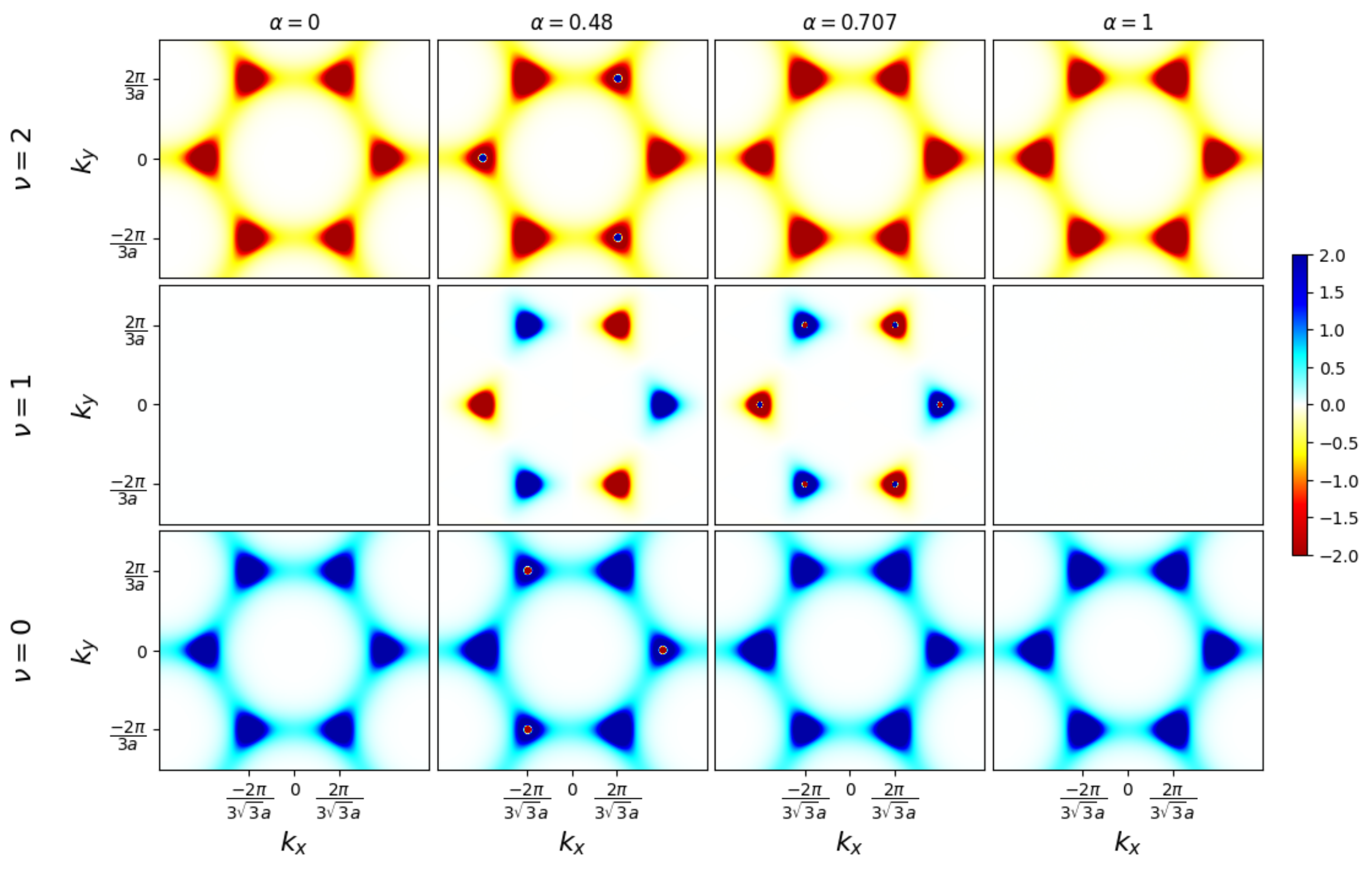}
	\caption{The Berry curvature distribution in the $k_{x}-k_{y}$ plane,
		corresponding to the conduction ($\protect \nu =0$), flat ($\protect \nu =1$)
		and valence ($\protect \nu =2$) bands, is calculated for different values of
		parameter $\protect \alpha $: $\protect \alpha =0$ (the graphene case), $%
		\protect \alpha =0.48$, $\protect \alpha =1/\protect \sqrt{2}$ (critical value
		corresponding to the phase transition where the band becomes dispersive at
		the Dirac point, as illustrated in Fig.\protect \ref{fig1}-(d)), and $\protect%
		\alpha =1$ (dice lattice limit). Calculations are performed by setting $%
		\protect \Delta =0.18\protect \gamma $ in the standard case where $\protect%
		\gamma _{1}=1\protect \gamma $, without any deformation.}
	\label{fig2}
\end{figure}

\begin{figure}[H]
	\centering
	\includegraphics[width=0.78\linewidth]{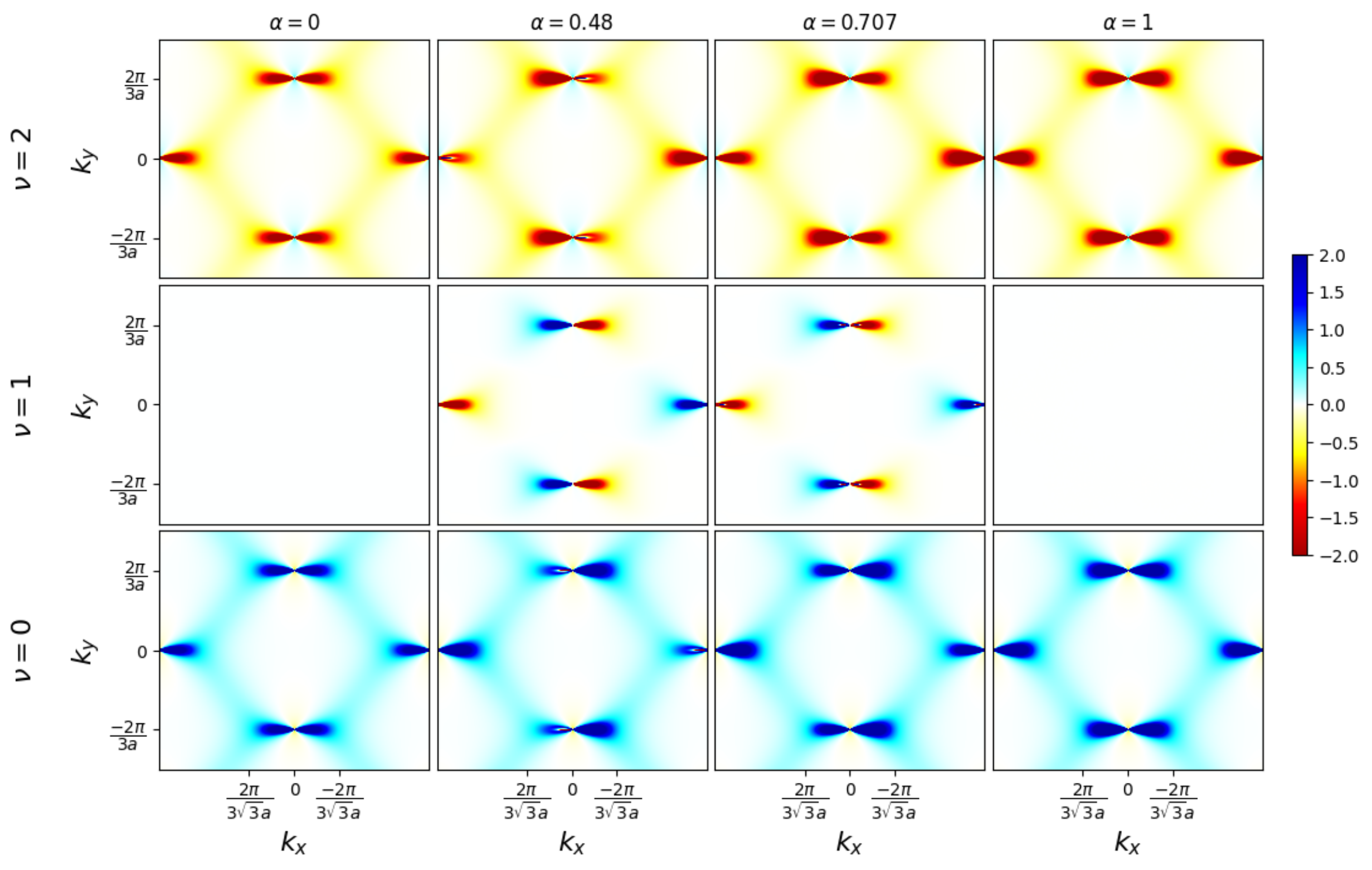}
	\caption{The Berry curvature distribution in the presence of deformation by
		modifying $\protect \gamma _{1}=2\protect \gamma $, while maintaining the
		other parameters in Fig.\protect \ref{fig2}.}
	\label{fig3}
\end{figure}

In this subsection we analyze the Berry curvature, which
plays a crucial role in topological quantum physics. It is characterized by
the operations of the following discrete symmetries: time inversion symmetry 
$\hat{\mathcal{T}}^{-1}\varOmega_{\nu }(\bm k)\hat{\mathcal{T}}=-\varOmega%
_{\nu }(-\bm k)$, charge conjugation symmetry $\hat{\mathcal{C}}^{-1}%
\varOmega_{\nu }(\bm k)\hat{\mathcal{C}}=-\varOmega_{\bar{\nu}}(-\bm k)$,
and inversion symmetry $\hat{\mathcal{I}}^{-1}\varOmega_{\nu }(\bm k)\hat{%
\mathcal{I}}=\varOmega_{\nu }(-\bm k)$, where $\bar{\nu}$ represents the
quasi-energy index conjugate of $\nu $ ($\varepsilon _{\bar{\nu}}(\bm %
k)=-\varepsilon _{\nu }(\bm k)$). We are interested in the effect of the
variation of $\gamma _{1}$ on this Berry curvature symmetry. To do this, we
numerically calculate the Berry curvature of the system in the z-component,
which is given by \cite{M7} 
\begin{equation}
\varOmega_{\nu }(\bm k)=-2\mathfrak{Im}\sum_{\nu ^{\prime }\neq \nu }\dfrac{%
\bra{\Upsilon_{\nu}(\bm k)}v_{x}\ket{\Upsilon_{\nu'}(\bm k)}%
\bra{\Upsilon_{\nu'}(\bm k)}v_{y}\ket{\Upsilon_{\nu}(\bm k)}}{(\varepsilon
_{\nu }(\bm k)-\varepsilon _{\nu ^{\prime }}(\bm k))^{2}},  \label{eq15}
\end{equation}%
where $v_{i}=\hbar ^{-1}\partial _{k_{i}}H(\bm k)$ is the effective velocity
in the axial direction $i=x,y$. We know that the Berry curvature, if not
zero, results from the breaking of the time-reversal symmetry, or at least
from the presence of a broken symmetry, as mentioned previously. We now plot
the Berry curvature to analyze this symmetry and to study the effect of
variations of $\gamma _{1}$. We vary $\gamma _{1}$ in two cases: $\gamma
_{1}=\gamma $ and $\gamma _{1}=2\gamma $.\newline
For $\gamma _{1}=\gamma $, time-reversal symmetry is broken when Berry
curvatures differ from zero for any value of $\alpha$ and any individual Berry
curvature. Charge conjugate symmetry is present for $\alpha =0$ and 1, see
Fig .\ref{fig2} where the Berry curvature $\hat{\mathcal{C}}^{-1}\varOmega%
_{2}(\bm k)\hat{\mathcal{C}}=-\varOmega_{0}(-\bm k)$ and the associated
curvature of a flat band $\varOmega_{1}$ is zero, and when k is given, it's
clear that the sum of the individual Berry curvatures is zero, so the local
conservation of Berry curvature and also the inversion symmetry is present $%
\hat{\mathcal{I}}^{-1}\varOmega_{2}(\bm k)\hat{\mathcal{I}}=\varOmega_{2}(-%
\bm k)$ and $\hat{\mathcal{I}}^{-1}\varOmega_{0}(\bm k)\hat{\mathcal{I}}=%
\varOmega_{0}(-\bm k)$, but for $\alpha \neq 0$ and 1 the conjugate charge
and inversion symmetry are broken. The expression for the Berry curvature
can be found analytically in the case where $\alpha =1$, and is written like
this 
\begin{equation}
\varOmega_{\nu }(\bm k)=\dfrac{\Theta (\bm k)}{(|\rho (\bm k)|^{2}+\eta ^{2}(%
\bm k)/2)^{3/2}}(\delta _{\nu ,0}+0\ast \delta _{\nu ,1}-\delta _{\nu ,2}),
\end{equation}%
and 
\begin{equation}
\begin{split}
\Theta (\bm k)=& \frac{\sqrt{3}}{8\gamma }a^{2}\Delta \bigg(3\gamma (\gamma
^{2}+2\gamma _{1}^{2})+4\gamma (\gamma ^{2}-2\gamma _{1}^{2})\cos (\sqrt{3}%
ak_{x})+\gamma ^{3}\cos (2\sqrt{3}ak_{x}) \\
& +\gamma _{1}\big[2(-3\gamma ^{2}+\gamma _{1}^{2})\cos \left( \frac{1}{2}a(%
\sqrt{3}k_{x}-3k_{y})\right) +\gamma \gamma _{1}\cos \left( a(\sqrt{3}%
k_{x}-3k_{y})\right)  \\
& +2(-3\gamma ^{2}+\gamma _{1}^{2})\cos \left( \frac{1}{2}a(\sqrt{3}%
k_{x}+3k_{y})\right) +\gamma \gamma _{1}\cos \left( a(\sqrt{3}%
k_{x}+3k_{y})\right) \big]\bigg).
\end{split}%
\end{equation}%
As we saw previously, the Berry curvature is concentrated at Dirac points $%
\bm K$ and $\bm K^{\prime }$ for $\gamma _{1}=\gamma $ in the Brillouin zone
(BZ). As $\gamma _{1}$ increases, we observe that the Berry curvature moves
slowly towards point $\bm M$ for $\gamma _{1}=2\gamma $ (see Fig.\ref{fig3}%
). However, this figure does not include results for other values of $\gamma
_{1}$. The Berry curvature becomes singular at $\gamma_{1}=2\gamma $, where
quasi-energy is degenerate at point $\bm M$. When we vary $\gamma _{1}$, the
system breaks the inversion symmetry and, more precisely, the $C_{3}$
symmetry is also broken in the cases where $\alpha =0$ and $\alpha =1$ (see
Fig.\ref{fig3}).
\begin{figure}[H]
	\centering
	\includegraphics[width=0.92\linewidth]{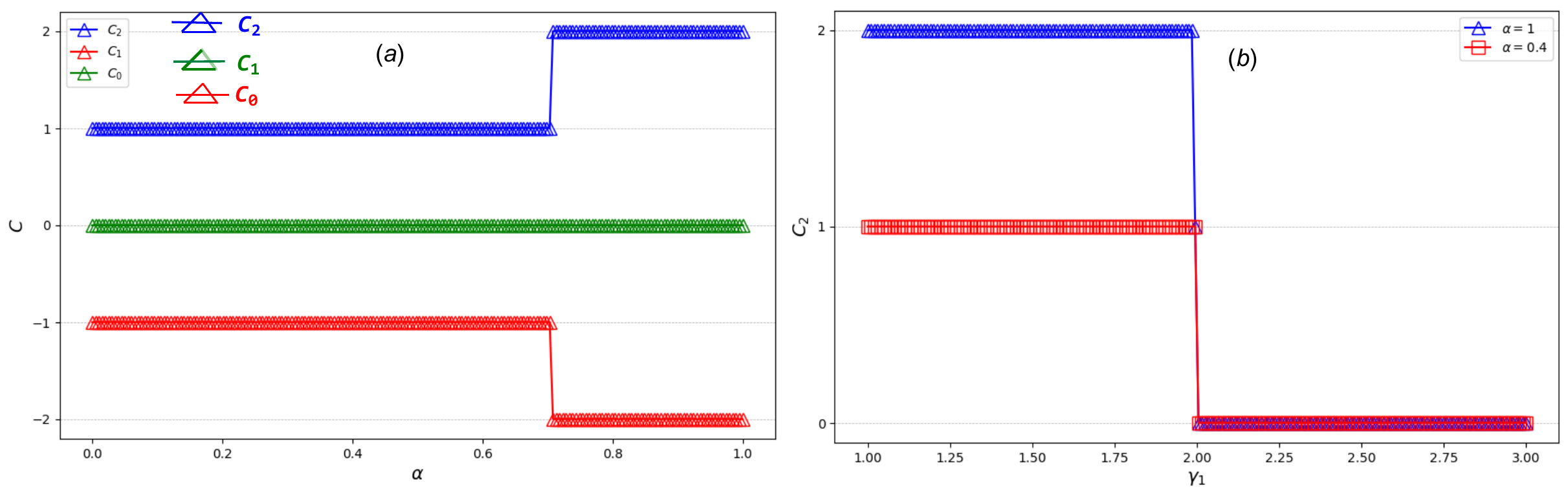}
	\caption{The variation of the Chern number as a function of the parameters $%
		\protect \alpha $ and $\protect \gamma _{1}$: For (a) The variation of the
		Chern number corresponding to the valence ($\protect \nu =2$), flat ($\protect%
		\nu =1$) and conduction ($\protect \nu =0$) bands is shown as a function of $%
		\protect \alpha $, as illustrated in the figure above. In (b), the variation
		of the Chern number corresponding to the valence band ($\protect \nu =2$) as
		a function of $\protect \gamma _{1}$ is presented. This represents the
		deformation of the system for $\protect \alpha =0.4$ and $\protect \alpha =1$.
		In all the calculations presented here, we have fixed the parameter $\Delta
		=0.18\protect \gamma $.}
	\label{fig4}
\end{figure}
\subsection{Chern phase diagrams}\label{SubsecB}
\begin{figure}[H]
	\centering
	\includegraphics[width=1\linewidth]{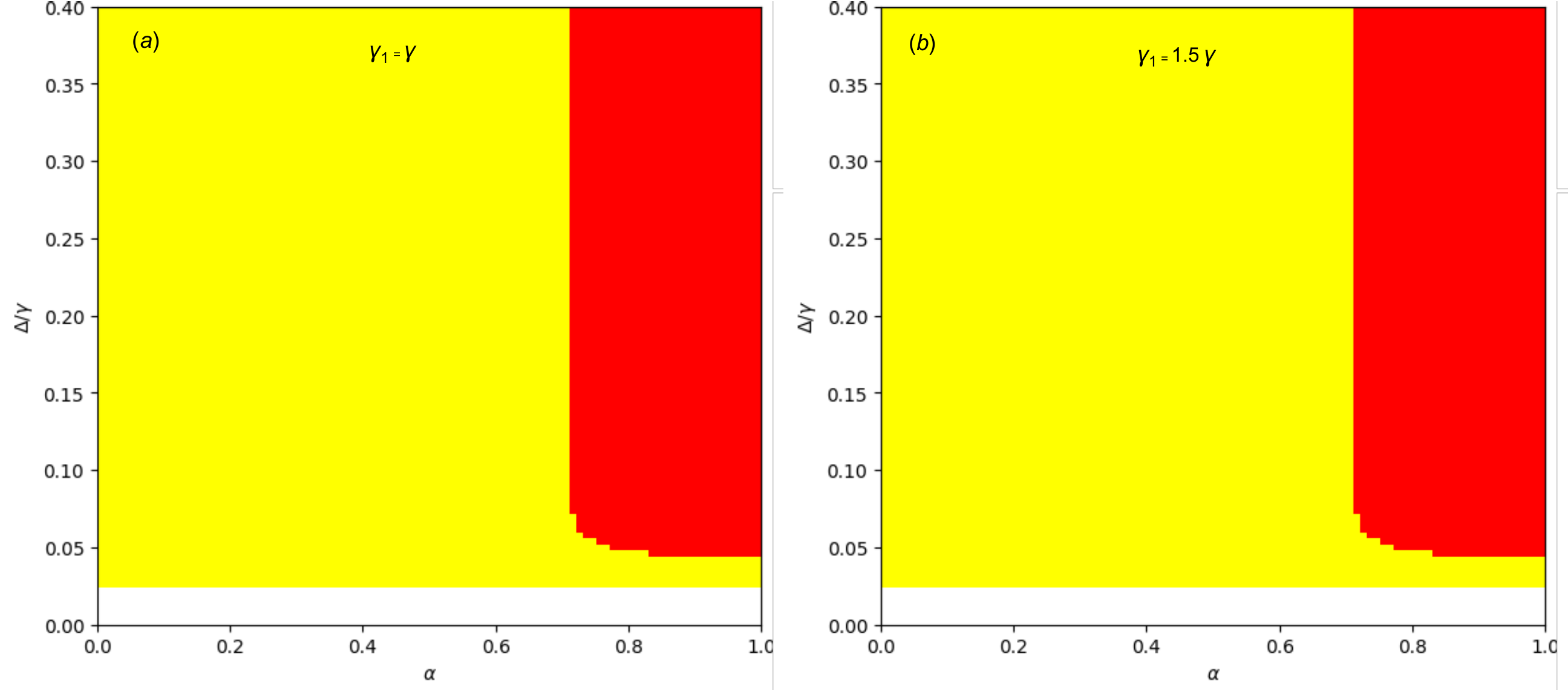}
	\caption{The phase diagram of the Chern number associated with the valence
		band ($\protect \nu =2$) of the irradiated and deformed $\protect \alpha -T_{3}
		$ lattice is presented in parameter space ($\Delta /\protect \gamma \in $[0
		0.4]) and ($\protect \alpha \in $[0 1]). The phase diagram is shown for (a) $
		\protect \gamma _{1}=1\protect \gamma $ and (b) $\protect \gamma _{1}=1.5\protect\gamma $  .}
	\label{fig5}
\end{figure}

 In this subsection, we aim to obtain the phase diagram of
the Chern number. As we have previously mentioned, light induces a term
similar to Haldane's term for $\phi =\dfrac{\pi }{2}$, which is responsible
for breaking the time-reversal symmetry. This term, as we have studied,
generates a non-zero Berry curvature, which is concentrated at the Dirac
points due to the opening of a gap caused by light. This implies the
existence of a phase transition, which can be observed through the surface
integral of the Berry curvature over the BZ. This integral gives an integer,
called the Chern number for the $\nu $-th band, which is defined by the
following relationship \cite{M7}. 
\begin{equation}
C_{\nu }=\dfrac{1}{2\pi }\int \int_{BZ}\varOmega_{\nu }(\bm k)d^{2}\bm k.
\end{equation}%
Before discussing the evolution of the Chern number as a function of
parameter variations, we first consider the standard case where no
deformation varies the hopping energy $\gamma _{1}$, i.e. $\gamma
_{1}=\gamma $. In the $\alpha -T_{3}$ lattice there are three bands: two
dispersive bands ($\varepsilon _{0}$ and $\varepsilon _{2}$) and a flat band
($\varepsilon _{1}$), which becomes dispersive at the Dirac point when $%
\alpha =1/\sqrt{2}$. To observe the topological evolution of the system, we
study the transition between the graphene lattice ( $\alpha =0$) and the
dice lattice limit ($\alpha =1$), varying $\alpha $ from 0 to 1. The system
becomes topologically non-trivial due to a non-zero Chern number, except for
the Chern numbers associated with the bands $\nu =0$ and $\nu =2$. We see
that there is a phase transition at $\alpha =1/\sqrt{2}$, where the Chern
numbers of the bands are $C_{2}(C_{0})=1(-1)$ in the range $0\leq \alpha <1/%
\sqrt{2}$. At $\alpha =1/\sqrt{2}$ these values change from $%
C_{2}(C_{0})=1(-1)$ to $C_{2}(C_{0})=2(-2)$ when $1/\sqrt{2}\leqslant \alpha
\leqslant 1$ (see Fig.\ref{fig4}-(a)). On the other hand, the Chern number
associated with the band $\nu =1$ remains zero, indicating that it is
topologically trivial. 
We now consider the $\alpha -T_{3}$ lattice irradiated and deformed along
the displacement vector $\mathbf{\delta }_{1}$, resulting in a change in the
hopping energy $\gamma _{1}$. For this study we choose a variation region of 
$\gamma _{1}$ between $[\gamma ,3\gamma ]$. We investigate the evolution of
the Chern number $C_{2}$, which is the inverse of $C_{0}$, while $C_{1}$
remains topologically trivial. Fixing $\Delta =0.18\gamma $ and $\alpha =0.4$
as well as $\gamma =1$, we plot $C_{2}$ as a function of $\gamma _{1}$. For $%
\gamma _{1}=\gamma$, the Chern number $C_{2}$ is initially equal to 1 for $%
\alpha =0.4$ and 2 for $\alpha =1$, as shown in Fig.\ref{fig4}-(a). As $%
\gamma _{1}$ increases from g to $\gamma _{1}<2\gamma $, $C_{2}$ remains
constant regardless of the value of $\alpha $. However, as $\gamma _{1}$
exceeds $2\gamma $, the system undergoes a change in topological properties
from a topologically non-trivial to a topologically trivial state. This
change, characteristic of a phase transition, occurs precisely at $\gamma
_{1}=2\gamma $. As shown in Fig.\ref{fig4}-(b), the Chern number $C_{2}$
changes from 1 ($\alpha =0.4$) or 2 ($\alpha =1$) for $\gamma _{1}=\gamma $
to 0 ($\alpha =0.4$) or 0 ($\alpha =1$) for $\gamma _{1}>2\gamma $. We also analyse variations in the Chern number $C_{2}$ as a function of the parameters $\Delta /\gamma $ and $\alpha $. The value of $\gamma _{1}$ is set to two distinct cases: $\gamma _{1}= 1\gamma$ and $1.5\gamma$. These cases are represented in the topological phase diagram in Fig. \ref{fig5}. This diagram uses a colour code to highlight the different topological phases of the system: red indicates a phase with $C_{2}=2$, yellow corresponds to $C_{2}=1$, and white designates a topologically trivial phase characterised by $C_{2}=0$. When $\gamma _{1}<2\gamma$ (see sub-figs \ref{fig5}(a) and \ref{fig5}(b)), we observe that, for low values of the polarisation amplitude ($\Delta /\gamma \in \lbrack 0,0.025]$), the system remains in a topologically trivial phase with $C_{2}=0$. However, as soon as $\Delta /\gamma > 0.025$, a topological phase transition occurs: the Chern number changes from $C_{2}=0$ to $C_{2}=1$ in the interval $\alpha \in\lbrack 0,1/\sqrt{2}]$. Furthermore, for $\Delta /\gamma =0.05$, a second transition is observed in the interval $\alpha \in \lbrack 0.85,1]$, where $C_{2}$ evolves from 1 to 2. On the other hand, when $\gamma _{1} > 2\gamma $, the system adopts a topologically trivial phase characterised by $C_{2}= 0$, irrespective of the values of $\Delta /\gamma $ and $\alpha$ (the diagrams corresponding to this case have not been included here). This behaviour is consistent with the results presented earlier in Fig. \ref{fig4}(b), where a critical transition to a topologically trivial insulator was identified when $\gamma _{1}$ exceeds the threshold value of 2$\gamma$. \newline
To confirm whether or not the system remains in a non-trivial topological
insulating state as $\gamma_{1}$ varies, we would like to study the
evolution of the Wannier charge center in the next subsection.
\subsection{Wannier charge centers}\label{SubsecC} 
This subsection focuses on examining Wannier charge
centres(WCCs) as a means to observe and analyse topological evolution, a
fundamental aspect of the diagnostic process of topological band properties 
\cite{M9,M11}. These centres are defined as the mean charge position in a
system based on Wannier functions. This mathematical conceptualisation finds
application in $\nu $ bands of a two-dimensional system, defined as follows 
\cite{M10,M11} 
\begin{equation}
\psi _{\nu }=\dfrac{ia}{2\pi }\int_{-\pi /a}^{\pi /a}\bra{\Upsilon_{\nu}(\bm
k)}\nabla _{\bm k}\ket{\Upsilon_{\nu}(\bm k)}d\bm k.  \label{eqwccs}
\end{equation}%
According to modern polarisation theory \cite{M9,M10} it is well established that WCCs are equivalent to Berry phase calculations.  However, to analyse the topological evolution of the system and extract relevant information, WCCs must be calculated along the $k_x$ direction as a function of the $k_y$ transverse momentum.  This method is commonly employed to study the topological invariant $Z_{2}$ and edge states in $\alpha-T_{3}$ lattices, particularly about the transition to the spin quantum Hall effect \cite{wwc2024}. This calculation is based on a modified version of equation (\ref{eqwccs}). 
\begin{equation}
\psi _{\nu }(k_{y})=\dfrac{ia}{2\pi }\int_{-\pi /a}^{\pi /a}%
\bra{\Upsilon_{\nu}( k)}\partial _{k_{x}}\ket{\Upsilon_{\nu}( k)}dk_{x}.
\label{eq20}
\end{equation}%
\begin{figure}[H]
	\centering
	\includegraphics[width=1\linewidth]{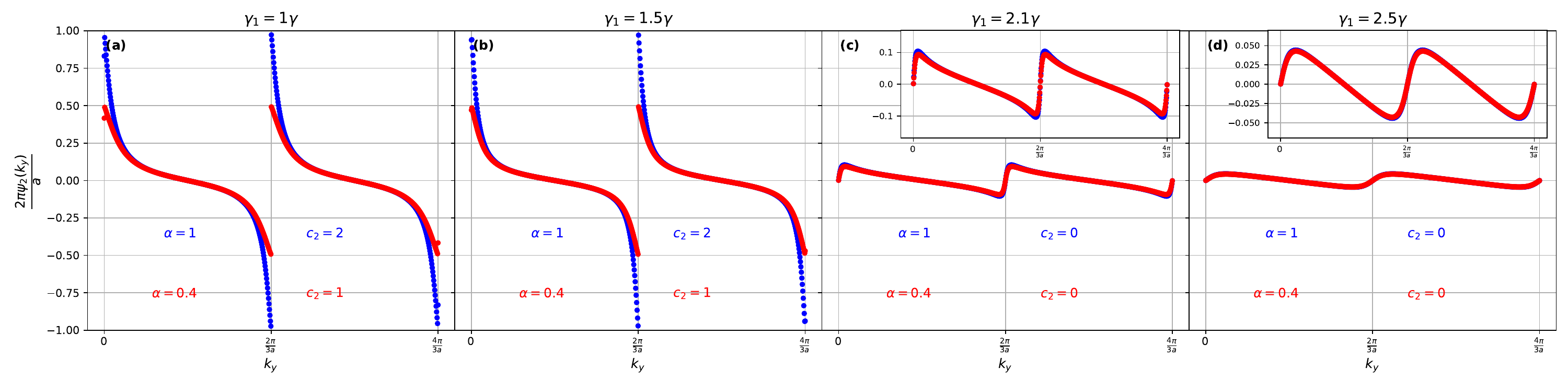}
	\caption{Evolution of the Wannier charge center for the valence band ($%
		\protect \nu =2$) along the $x$ direction as a function of displacement $k_{y}
		$ in the $y$ direction. For (a) $\protect \gamma _{1}=1\protect \gamma $ and
		(b) $\protect \gamma _{1}=1.5\protect \gamma $ , this evolution is
		characterized by a winding and discontinuity of the WCC at the Dirac points
		along $k_{y}$, when a closed cycle is traversed in the BZ, indicating a
		non-trivial topological phase. On the other hand, for (c) $\protect \gamma %
		_{1}=2.1\protect \gamma $ and (d) $\protect \gamma _{1}=2.5\protect \gamma $,
		the winding of the WCCs disappears, the latter becoming continuous, meaning
		that the system is no longer in a topological insulating phase.  The other parameters taken into account are $\protect \alpha =0.4$, corresponding to the WWC curve shown in red, $\protect \alpha = 1$, corresponding to the WWC curve shown in blue, and $\Delta =0.18\protect\gamma $.}
	\label{fig6}
\end{figure}

From the calculation of the Chern number associated with the valence bands, we can see that it becomes non-zero when $\gamma_{1}<2\gamma$. For example, when $\alpha= 1$, $C_2=2$, whereas when $\alpha= 0.4$, $C_2=1$. Conversely, annihilation occurs when $\gamma_{1}>2\gamma$, resulting in a Chern number of zero ($C_2=0$). The Chern number can be interpreted as the number of windings of the WCCs as they evolve with respect to $k_y$ in a one-dimensional system \cite{wwc2015}. In this context, we consider the busy case with $\nu = 2$ for $\alpha=1$ and $\alpha=0.4$. The WCCs are calculated along the $k_x$ direction, corresponding to one-dimensional integration over a closed loop. Their evolution is then studied as a function of $k_y$, in accordance with equation (\ref{eq20}). This analysis is performed for $k_y$ belonging to the intervals $[0, 2\pi]$, representing the displacement of the Wannier charge centre through a closed path in the first Brillouin zone. Charge conservation in the presence of edge states crossing the band gap \cite{In55} dictates the movement of charges, allowing transfer between the valence band and the conduction band. The WCC function, denoted $2\pi\psi _{2}(k_{y})/a$, can be represented graphically, as shown in Fig.\ref{fig6}, for $\alpha = 1$ and $\alpha = 0.4$.  We vary the hop parameter $\gamma_{1}$ for four distinct values: two in the $\gamma_{1}<2\gamma$ regime ($\gamma_{1}=1\gamma$, $\gamma_{1}=1.5\gamma$) and two in the $\gamma_{1}>2\gamma$ regime ($\gamma_{1}=2.1\gamma$, $\gamma_{1}=2.5\gamma$). The function $2\pi\psi _{2}(k_{y})/a$, assimilated to an angle (i.e., the Berry phase) \cite{wwc2015}, is then represented on a cylindrical surface in cylindrical coordinates, with $k_y$ in the longitudinal direction. When $\gamma_{1}<2\gamma$, two complete windings of the WCCs are observed for $\alpha=1$, corresponding to $C_2= 2$. For $\alpha=0.4$, however, the evolution of the WCCs does not appear to form a complete winding. However, the topological calculation indicates a Chern number $C_2= 1$. This means that even if the visual trajectory appears incomplete, all branches of the WCCs connect globally to form a winding, reflecting an effective winding and therefore a non-zero topological invariant ($C_2= 1$).  As shown in sub-figures \ref{fig6}(a) and \ref{fig6}(b), a discontinuity appears at the Dirac points in the evolution of the WCCs for $\alpha = 1$ and $\alpha = 0.4$. However, this discontinuity actually corresponds to a continuous transition on the opposite side of the cylinder. Conversely, no winding is observed on the cylindrical surface for $\gamma_{1}>2\gamma$. The Chern number then becomes zero ($C_2= 0$) for $\alpha = 1$ and $\alpha = 0.4$, as illustrated in sub-figures \ref{fig6}(c) and \ref{fig6}(d). Thus, the evolution of the WCCs reveals a topological transition: the system shifts from a non-trivial state ($\gamma_{1}<2\gamma$), characterised by WCC windings, to a trivial state ($\gamma_{1}>2\gamma$), where the WCCs stop to wind and oscillate around zero.

\section{Anomalous Hall Conductivity}\label{secV} 
\begin{figure}[H]
	\centering
	\includegraphics[width=0.8\linewidth]{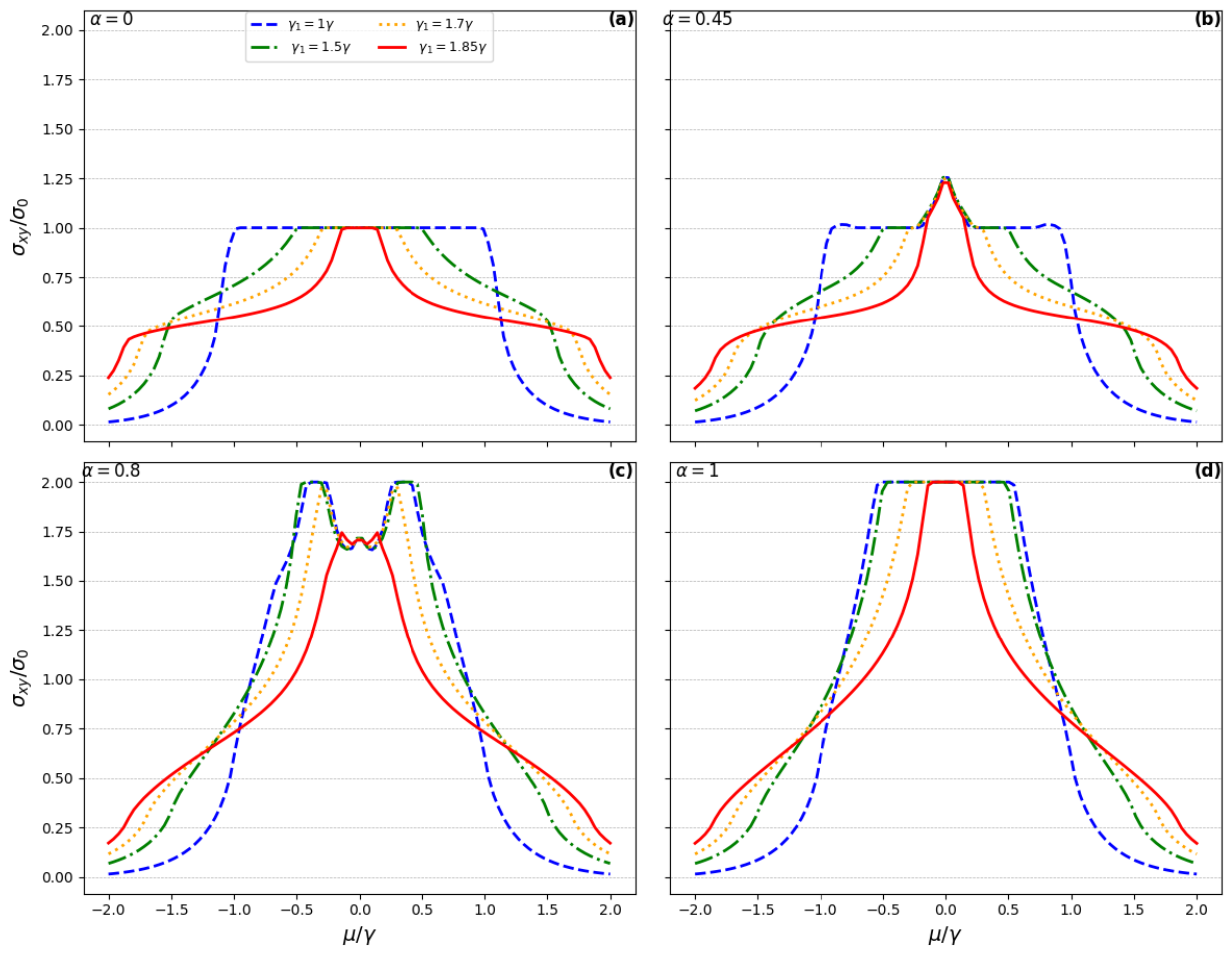}
	\caption{Hall conductivity is depicted as a function of chemical potential $%
		\protect \mu $ for different values of $\protect \gamma _{1}$, as shown in the
		inset. The cases studied are (a) $\protect \alpha =0$, (b) $\protect \alpha %
		=0.4$, (c) $\protect \alpha =0.8$ and (d) $\protect \alpha =1$.}
	\label{fig7}
\end{figure}

In this section, we calculate and discuss the anomalous Hall
conductivity (AHC) for the $\alpha -T_{3}$ lattice exposed to irradiation and deformation. This conductivity is evaluated by examining the electron response to an external electric field in an anisotropic context, providing information about the system topology. To determine the AHC, we integrate the Berry curvature over all occupied states in the entire Brillouin zone (BZ), by Eq.(\ref{eq15}). We use the following formula to perform this
calculation \cite{M7}. 
\begin{equation*}
\sigma _{xy}=\frac{\sigma _{0}}{2\pi }\sum_{\nu }\int_{BZ}d^{2}k\, \varOmega%
_{\nu }(\bm k)f_{\nu }(\bm k).
\end{equation*}%
Where $f_{\nu }(\bm k)=\left[ 1+e^{(\varepsilon _{\nu }(\mathbf{k})-\mu
)/k_{B}T}\right] ^{-1}$ is the Fermi-Dirac distribution function, $\mu $ is the chemical potential, T is the temperature, $k_{B}$ is the Boltzmann constant and $\sigma _{0}=e^{2}/h$. The AHC can be calculated numerically as
a function of the chemical potential $\mu $ at T=100 K by fixing $\Delta=0.35\gamma $ and $\alpha $ for four cases: $\alpha =0,0.45,0.8$ and 1, while varying $\gamma _{1}$, as shown in Fig.\ref{fig7}.\newline 
We first examine the first two cases, $\alpha =0$ and $\alpha =1$, where the hopping energy $\gamma _{1}$ varies, as shown in Fig. \ref{fig7}(a) and (d). A quantised plateau is observed with $\sigma _{xy}=\sigma _{0}$ for $\alpha =0$ and $\sigma _{xy}=2\sigma _{0}$ for $\alpha =1$. When the chemical potential $\mu $ lies within the global gap, the width of the plateau corresponds to that of the gap in the dispersion spectrum and increases proportionally with increasing $\Delta $. It is noteworthy that $\sigma _{xy}$ for $\alpha =1$ is twice that for $\alpha =0$. In this case, the Berry curvature associated with the flat band ($\nu =1$) disappears and has no
contribution to $\sigma _{xy}$. When $\mu $ intercepts the bands (whetherconduction $\varepsilon _{0}$ or valence $\varepsilon _{2}$), $\sigma _{xy}$ decreases. This decrease is explained by the fact that the integral isperformed on occupied states and $\mu $ has left the global band.
We observe conductivity comparable to that of the Haldane model when the field amplitude is controlled with $\Delta=t_{2}$,
$\alpha=1$ and flux $\phi=\pi/2$ \cite{In47}. Furthermore, this conductivity value is close to the sum of the Hall conductivities from each valley in the system's low-energy regime undergoing a Floquet transition\cite{In43}.\newline
In addition, the width of the plateau decreases with increasing $\gamma _{1}$, as the overall gap between the dispersive bands narrows. At $\gamma_{1}=2\gamma $ the plateau disappears completely, as does the Hall conductivity.\newline
In cases where $\alpha \neq0,1$, the flat band becomes dispersive, significantly modifying the behaviour of the Hall conductivity. This dispersive nature leads to smoother plateaus appearing in the Hall conductivity, as evident when $\alpha=0.45$. At this value, the flat band acquires a dispersion that creates an inhomogeneous gap at the $\bm K$ and  $\bm K'$ points, resulting in a finite Berry curvature. This reduces the overall contribution of all occupied states. In this scenario, two distinct $\sigma_{xy}$ plateaus are observed: one when the Fermi level $\mu$ lies between the $\varepsilon _{1 }(\bm k)$ and $\varepsilon _{2 }(\bm k)$ bands at the $\bm K$ point and one when it lies between the $\varepsilon _{0 }(\bm k)$ and $\varepsilon _{1 }(\bm k)$ bands at the $\bm K'$ point. The value of $\sigma_{xy}$ on each of these plateaus is equal to $\sigma_{0}$. This corresponds to the sum of the topological contributions of each low-energy valley in the $\alpha-T_{3}$ lattice, giving an approximate conductivity of $\sigma_{0}$ in the transition regime induced by Floquet-type irradiation \cite{In43}. However, rather than a well-quantised plateau, a $\sigma_{xy}$ summit emerges near $\mu = 0$. At this value of $\alpha = 0.45$, the flat band exhibits dispersive behaviour, resulting in a non-zero Berry curvature. Integrating this curvature yields partial contributions from several topological bands close to the Fermi level. While these partially occupied bands contribute to the Hall conductivity, they do not result in full quantisation. This behaviour can be explained by the dispersive nature of the flat band coupled with the absence of a sufficiently wide gap to isolate the contributions of each band. Furthermore, breaking charge conjugation symmetry can lead to Hall conductivity exhibiting an unquantised (i.e. non-integer) summit, as is observed here with a value of $\sigma_{xy}$ = 1.25$\sigma_{0}$.\newline
For $\alpha=0.8$, behaviour similar to that observed for $\alpha=0.45$ is evident: the flat band becomes dispersive, which has a significant impact on the topology of the energy spectrum. This dispersion results in two 
inhomogeneous gap openings, which are located at the $\bm K$ and $\bm K'$ points of the Brillouin zone. The flat band is dispersive, thus acquiring a non-zero Berry curvature, which reduces the overall contribution of the 
occupied states to the Hall conductivity. In this context, two quantised Hall conductivity plateaus appear, each with a value of $\sigma_{xy}= 2\sigma_{0}$. This increase in conductivity is the result of a topological 
transition that occurs at $\alpha= 1/\sqrt{2}$, as discussed previously. Indeed, the Chern number $|C_{0,2}|$ remains constant at 2 in the interval $1/\sqrt{2} < \alpha <1$, thus maintaining the value of $\sigma_{xy}$ at 
$\sigma_{0}$. This value is close to the conductivity summation for each 
valley \cite{In43}. These two flat sections appear when the $\mu$ is between the two 
bands that have some gaps, similar to what happens when $\alpha=0.45$.
In this regime, however, the Hall conductivity does not exhibit well-defined plateaus near the $\mu=0$ level. Instead, a significant decrease in  $\sigma_{xy}$ is observed at this point, which can be attributed to the non-negligible contribution of the dispersive flat band. The associated Berry curvature induces partial integration, resulting in several topologically active bands close to the Fermi level. As these bands are only partially occupied, full $\sigma_{xy}$ quantisation is not possible, resulting in a reduction of the Hall conductivity around $\mu=0$, as shown in sub-figure (c).\newline
Here, for $\alpha =0.45$ and $\alpha =0.8$, the width of the plateaus in $\sigma _{xy}$ decreases with increasing $\gamma_{1}$ as long as $\gamma_{1}<2\gamma $ and disappears completely when $\gamma _{1}\geq 2\gamma $.
\newline
To situate our findings within the broader context of recent research on the $\alpha-T_{3}$ lattice, we note that several important studies have explored Hall conductivity and topological phase transitions under mass terms, Zeeman coupling, and Floquet driving, primarily using low-energy continuum models near the Dirac points \cite{In43,AHCad1,AHCad2,AHCad3,AHCad4,AHCad5}. While thematically related, our work presents a distinct and complementary approach. First, we employ a full tight-binding model, allowing access to both low- and high-energy features across the entire Brillouin zone, essential for precise calculations of the Chern number and Hall conductivity. Second, whereas prior works typically address either mass perturbations or Floquet effects separately, we investigate their combined influence—namely, the interplay between anisotropic hopping ($\gamma_{1}\neq\gamma$) and circularly polarized light—revealing new and rich topological phase structures. Third, our analysis focuses on the total anomalous Hall conductivity across the full three-band system, rather than valley- or spin-resolved components, thereby capturing global topological features. Additionally, we introduce an explicit Wannier charge center (WCC) analysis, whose winding behavior provides a real-space signature of topological transitions—an aspect not addressed in the aforementioned references. Fourth, our multidimensional phase diagrams, constructed as functions of both deformation and light parameters, provide a broader and more tunable perspective on phase control in the $\alpha-T_{3}$ lattice. Finally, we find that the value of $\alpha$ plays a crucial role in increasing the Hall conductivity as well as in widening the corresponding plateaus. These plateaus are proportional to the Chern number, with $\sigma_{xy}=|C_{0,2}|\sigma_{0}$ as long as the system remains a non-trivial topological insulator. However, these plateaus are sensitive to the hopping energy $\gamma_{1}$, which leads to a progressive reduction in their width as $\gamma_{1}$ increases. Ultimately, these plateaus disappear when the band gap closes at the critical point $\gamma_{1} = 2\gamma$, marking the system's transition to a trivial topological insulator state.

\section{Conclusion}\label{secVI}
 In this work, we have investigated the effect of deformation
on the topological properties induced by light polarisation in $\alpha -T_{3}$ lattices. We introduced this deformation by modifying the hopping energy in the $\alpha -T_{3}$ lattice, particularly by modifying $\gamma _{1}$ at the A-B and A-C sites along the $\delta _{1}$ direction, while remaining unchanged hopping energies along the $\delta _{2}$ and $\delta _{3}$
directions. This modification of $\gamma _{1}$ led to a change in the band structure for three specific cases: $\alpha =0$, $\alpha =1/\sqrt{2}$ and $\alpha =1$. In these cases, the Dirac cones at points $\bm K$ and $\bm K^{\prime }$ move towards point $\bm M$ when the system is subjected to off-resonance circular polarization. This polarisation induces a Haldane mass term at $\phi =\pi /2$, breaking the time-reversal symmetry. This term
is responsible for the opening of a gap. However, in the case of $\alpha =1/\sqrt{2}$, the gap partially opens at points $\bm K$ and $\bm K^{\prime }$. Moreover, the gap size decreases with increasing $\gamma _{1}$ and disappears completely when $\gamma _{1}=2\gamma $, where the system adopts semi-Dirac behaviour. This deformation, which breaks the $C_{3}$ symmetry,leads to a shift in the concentration of Berry curvature in the Brillouin
zone , from points $\bm K$ and $\bm K^{\prime }$ to point $\bm M$. In the standard case, $\gamma =\gamma _{1}$, we observe particle-hole symmetry for $\alpha =0$ and $\alpha =1$ and inversion symmetry. We have calculated the corresponding Chern numbers for the conduction ($C_{0}$), flat ($C_{1}$) and
valence ($C_{2}$) bands: For $\alpha <1/\sqrt{2}$ the Chern numbers are $ C_{0}=-1$, $C_{1}=0$ and $C_{2}=1$. For $\alpha \geq 1/\sqrt{2}$ the values change to $C_{0}=-2$, $C_{1}=0$ and $C_{2}=2$. This shows a phase transition. We then focused on the Chern number corresponding to $C_{2}$. Fixing $\alpha =1$, we plotted $C_{2}$ as a function of $\gamma _{1}$. It
turned out that there is a phase transition with gap closure at $\gamma_{1}=2\gamma $. When $\gamma _{1}>2\gamma $, the system changes from a non-trivial topological insulator ($C_{2}=2$) to a trivial topological insulator ($C_{2}=0$). We have also plotted a phase diagram in the parameter space of $\Delta /\gamma $ and $\alpha $ by fixing $\gamma _{1}$. When $
\gamma _{1}>2\gamma $, in the interval of $\Delta /\gamma _{1}$ and $\alpha $, the system transforms into a trivial topological insulator. This transition was confirmed by plotting the Wannier charge centre. Fixing $\gamma _{1}$, we observed that the system changes state from a non-trivial topological insulator to another trivial topological insulator state when $\gamma _{1}>2\gamma $. We have also calculated the anomalous Hall
conductivity for four values of $\alpha $: 0, 0.45, 0.8, and 1. For $\alpha=0$ and $\alpha =0.45$, the conductivity shows quantized behaviour in the form of plateaus with a value of $\sigma _{xy}=1\sigma _{0}$. For $\alpha=0.8$ and $\alpha =1$, the conductivity value increases to $\sigma_{xy}=2\sigma _{0}$, characterized by a plateau at $\gamma _{1}=\gamma $.
This plateau decreases with increasing $\gamma _{1}$ and disappears when $\gamma _{1}>2\gamma $. These plateaus correspond to the Chern number, where $\sigma _{xy}=|C_{0,2}|\sigma _{0}$. The disappearance of these plateaus
indicates that the system is transitioning from a non-trivial topological insulator to a trivial topological state.\newline

\par In the context of the $\alpha-T_{3}$ lattice, the $\gamma_{1}$ parameter controls the uniaxial deformation of the lattice along a specific direction (e.g. the $\bm\delta_{1}$ and -$\bm\delta_{1}$ direction). The physical implications of this critical condition, $\gamma_{1}=2\gamma$, are significant. As $\gamma_{1}$ increases, the lattice is compressed in this direction, bringing the associated atomic sites closer together in the real lattice. When this compression reaches the critical value of  $\gamma_{1}=2\gamma$, the system enters a structural limit state in which sublattices overlap in real space due to inter-site distances becoming extremely small. This point corresponds to a significant geometric and electronic transition, during which the system's band properties and topology undergo radical alteration. Several signatures characterise this critical point, including the breaking of $C_{3}$ rotational symmetry. For $\alpha=1$, three-way crystal symmetry ($C_{3}$ rotation) is preserved when $\gamma_{1}=2\gamma$; however, it is broken when $\gamma_{1}\neq\gamma$, which results in the topological equilibrium between the $\bm K$ and $\bm K'$ points being disrupted.  At $\gamma_{1}=2\gamma$, the Dirac point moves into the Brillouin zone and merges with the $\bm M$ point. This results in a reconstruction of the energy spectrum. This merger can be interpreted as an overlap between sublattices in real space, which profoundly affects the system's Berry curvature. The Chern number, which characterises electronic band topology, changes from $C_{2}=2$ to $C_{2}=0$ when $\gamma_{1}>2\gamma$ for $\alpha=1$. This implies that all topological properties, such as protected edge states or quantised Hall conductivity, vanish beyond this threshold. The system then becomes topologically trivial and unable to maintain protected transport. From a physical point of view, $\gamma_{1}=2\gamma$ represents the maximum compression that the system can withstand without losing its topological coherence. Beyond this point, the lattice structure becomes too distorted to support a robust topological phase.
\section*{Acknowledgements}
The authors would like to acknowledge the " Hassan II Academy of Sciences and Techologies-Morocco for its financial support. The authors also thank the LPHE-MS, Faculty of Sciences, Mohammed V University in Rabat, Morocco for the technical support through facilities.

\end{document}